\newcommand\mdot{M$_{\odot}$~}
\def\spose#1{\hbox to 0pt{#1\hss}}
\newcommand\simlt{\mathrel{\spose{\lower 3pt\hbox{$\mathchar"218$}}
     \raise 2.0pt\hbox{$\mathchar"13C$}}}
\newcommand\simgt{\mathrel{\spose{\lower 3pt\hbox{$\mathchar"218$}}
     \raise 2.0pt\hbox{$\mathchar"13E$}}}
\newcommand{\civ}{\mbox{C$\,${\sc iv}$\lambda1549$}}
\newcommand{\mgii}{\mbox{Mg$\,${\sc ii}$\lambda2798$}}
\newcommand{\ciii}{\mbox{C$\,${\sc iii}]$\lambda1909$}}
\newcommand{\cii}{\mbox{C$\,${\sc ii}]$\lambda2326$}}
\newcommand{\oii}{\mbox{[O$\,${\sc ii}]$\lambda3727$}}
\newcommand{\oiii}{\mbox{[O$\,${\sc iii}]$\lambda\lambda4959,5007$}}
\newcommand{\nii}{\mbox{[N$\,${\sc ii}]$\lambda6583$}}
\newcommand{\hi}{\mbox{H$\,${\sc i}}}
\newcommand{\heii}{\mbox{He$\,${\sc ii}$\lambda1640$}}
\newcommand{\ion}[2]{\mbox{\textrm{#1}$\;$\textsc{#2}}}

\documentclass[useAMS,usenatbib]{mn2e}
\usepackage{psfig}
\usepackage{epsfig}
\usepackage{lscape}
\usepackage{amssymb}
\defcitealias{McC96}{McCarthy et al. (1996; 2007, private communication)}
\defcitealias{Bro07}{Paper\,I}
\defcitealias{Bry08}{Paper\,II}

\title[ Distant radio galaxies in the Southern hemisphere
-- III. ]{
A new search for distant radio galaxies in the Southern hemisphere -- III. Optical spectroscopy and analysis of the MRCR--SUMSS sample$\thanks{
Based on observations obtained at the ESO Very Large Telescope (VLT) and
New Technology Telescope (NTT) as part of program numbers 077.A-0471 and 079.A-0504 and the ANU 2.3\,m telescope.
}$}
\author[J. J. Bryant et al.]{J. J. Bryant$^{1}$\thanks{E-mail:
jbryant@physics.usyd.edu.au}, H. M. Johnston$^{1}$, J. W. Broderick$^{1}$,
R. W. Hunstead$^{1}$, C. De Breuck$^{2}$,\newauthor and B. M. Gaensler$^{1,3}$ 
\\
$^{1}$SIfA, School of Physics A29, The University of Sydney, NSW 2006, Australia \\
$^{2}$European Southern Observatory, Karl Schwarzschild Stra\ss e 2, D-85748 Garching, Germany\\
$^{3}$Harvard-Smithsonian Center for Astrophysics, 60 Garden Street, Cambridge MA 02138, USA}
\begin{document}

\date{}

\pagerange{\pageref{firstpage}--\pageref{lastpage}} \pubyear{2006}

\maketitle

\label{firstpage}

\begin{abstract}

We have compiled a sample of 234 ultra-steep-spectrum(USS)-selected radio
sources in order to find high-redshift radio
galaxies (HzRGs). 
The sample is in the southern sky at $-40^{\circ} < \delta < -30^{\circ}$ 
which is 
the overlap region of the 
408-MHz Revised Molonglo Reference Catalogue, 
843-MHz Sydney University Molonglo Sky Survey (the MRCR--SUMSS sample)
and the 
1400-MHz NRAO VLA Sky Survey. 
This is the third in a series of papers on the
MRCR--SUMSS sample. Here we present optical spectra from the 
ANU 2.3-m telescope, ESO New Technology
Telescope and ESO Very Large Telescope for 52 of the identifications from \citetalias{Bry08}, 
yielding redshifts
for 36 galaxies, 13 of which have $z>2$. We analyse the $K$--$z$ distribution and compare 
4-arcsec-aperture magnitudes with 64-kpc aperture magnitudes in several surveys from the
literature; the MRCR--SUMSS sample is found to be consistent with models for
$10^{11}$--$10^{12}$\,\mdot galaxies. 
Dispersions about the fits in the $K$--$z$ plot support passive evolution of radio galaxy
hosts since $z>3$.
By comparing USS-selected
samples in the literature, we find that the resultant median redshift of the
samples shown is not dependent on
the flux density distribution or selection frequency of each sample. In addition, our finding
that the majority of the radio spectral energy distributions remain straight over a wide frequency
range suggests that a k-correction is not responsible for the success of USS-selection in identifying
high redshift radio galaxies and therefore the steep radio spectra may be intrinsic to the source
or a product of the environment.
Two galaxies have been found to have both  
compact radio structures and strong self-absorption in the Ly$\alpha$ line, suggesting 
they are surrounded by a dense medium. 
For the bulk of the sources, spectral line ratios show that photoionisation 
is the primary excitation process.

\end{abstract}


\section{INTRODUCTION}
\label{Intro}

Hierarchical models of galaxy formation 
predict that massive galaxies formed through many mergers of
smaller systems, with continuous bursts of star formation. However,
there is solid evidence that the most massive galaxies in the
local Universe formed their stars very early in the history of the
Universe \citep{Sta:98,Nel:01,Tho:05}. In order to fit current
galaxy-formation models, feedback processes have been introduced in
which the energy input from a central massive black hole quenches the
star formation by heating the interstellar gas 
\citep{Spr:05,deL:05,Bow:06}. 
Such
modifications appear to overcome many of the shortcomings of previous
hierarchical models,
but although the
quenching of star formation is attributed to the onset of a galactic
wind driven by ionising radiation and/or mechanical energy from the
radio jets, the physical mechanisms for this are poorly understood.
In contrast, there is observational evidence
to suggest that the radio jets associated with some supermassive black
holes might {\em trigger\/}, rather than suppress, star formation in
high-redshift galaxies \citep{Ree:89,Bic00,Kla:04,Crof06}.  There are also
plausible arguments to suggest that both processes operate, with radio
jets first triggering star formation and then later driving a galactic
wind \citep{Sil:05}.
The interplay between mergers and feedback associated with active galactic 
nuclei (AGN) is likely to affect the
evolution of all massive galaxies, 
not just the extreme
bright end of the luminosity function \citep{Cro:06}. However, it is the most
massive galaxies at high redshift that provide the most stringent
constraints on these processes. 

The star formation rate density in the early Universe is dominated by the most massive
galaxies in the sample volume \citep[see fig.\,2 of][]{Feu05}. 
In \citet[][fig. 5]{Roc04}, radio galaxies have been
identified to be more massive than optically-selected galaxies
at each redshift, 
with stellar masses of $\sim 10^{12}\,{\rm M}_\odot$,
which is $5-10$ times more massive than the largest
galaxies in most optical surveys.
The star formation rate density 
of the Universe has been shown to peak at $2<z<3.5$ 
\citep[][fig.\,2]{Lil96,Mad96,Hop04,Hop06} 
and since radio 
galaxies are the most massive and dominate the star formation 
rate at these epochs, they are
crucial to understanding early galaxy evolution. 

We have conducted a survey of ultra-steep-spectrum (USS) radio sources in the Southern Hemisphere,
designed to find high-redshift ($z>2$) radio galaxies so as to carry out a
detailed quantitative study of their environments.
This is the first USS survey in the South to use large, deep low-frequency catalogues
to mine a large area of sky, and it became possible with the completion of
the Sydney University Molonglo Sky Survey at 843\,MHz 
\citep[SUMSS;][]{Boc99,Mau03} 
and the reanalysis of
the 408-MHz Molonglo Cross survey \citep{Lar81} 
to give the more sensitive revised Molonglo Reference Catalogue (MRCR; Crawford, private
communication).

The details of the sample selection, and the 1384- and 2368-MHz radio imaging of the
MRCR--SUMSS sample were
presented in \citet[][hereafter Paper\,I]{Bro07}.
The $K$-band imaging and identification of the counterparts to 175 of
the radio sources were presented in \citet[][hereafter Paper\,II]{Bry08}.
The current paper includes the followup spectroscopy for the sources in \citetalias{Bry08} and
a complete discussion of the sample using the radio, $K$-band, redshift and spectral
line information. A 
following paper will discuss the
environments of the $z>2$ galaxies and what that can tell us about the early
evolution of massive galaxies. 

We have adopted a flat, $\Lambda$ cold dark matter cosmology with  
$H_0=71$ km s$^{-1}$ Mpc$^{-1}$, $\Omega_{\rm M}=0.27$ and $\Omega_{\Lambda}=0.73$.

\section{Observations and data reduction}
\subsection{Target selection and radio and $K$-band imaging}

The target selection, radio observations and $K$-band imaging were discussed 
extensively in Papers I and II. A very brief summary is given here.

The MRCR--SUMSS sample was formed by cross-matching the 
revised
408-MHz Molonglo Reference Catalogue (MRCR), the 843-MHz SUMSS 
catalogue and the 1400-MHz NVSS catalogue \citep{Con98} in the region
where all three catalogues overlap (declination $-30^{\circ}$ to $-40^{\circ}$).
We then selected sources with a spectral index\footnote{Radio spectral index $\alpha$ 
is related to $S_{\nu}$, the flux density at frequency $\nu$, by 
$S_{\nu} \propto \nu^{\alpha}$} between 408 and 843\,MHz, $\alpha_{408}^{843}\leq -1.0$ and
Galactic latitude $|b|>10^{\circ}$. We further eliminated
any sources that had another NVSS source within 100\,arcsec 
to minimize the effects of source confusion.

Follow-up radio images were obtained with the Australia Telescope Compact Array (ATCA)
at 1384 and 2368\,MHz for the full sample,
to pinpoint the $K$-band identification and investigate the radio
structure. These are detailed in \citetalias{Bro07}.
Twenty-nine of the MRCR--SUMSS sources were also observed at 4800 and 8640\,MHz; those
data are presented in \citetalias{Bry08}.

$K$-band ($2.2\mu$m) observations were made with the 3.9-m Anglo-Australian 
Telescope's (AAT) IRIS2 detector \citep{gil00} 
and with PANIC \citep{mar04} 
on the 6.5-m Magellan Baade telescope at Las Campanas Observatory. 
Full details of the $K$-band imaging and reduction are given in \citetalias{Bry08}.

\subsection{Spectroscopy}

Spectra for our sources were obtained using three
spectrograph/telescope combinations. Sources with optical identifications from the
SuperCOSMOS Sky Survey \citep{Ham01} 
were observed using the 2.3-m telescope of the Australian National University (ANU) at
Siding Spring Observatory, NSW. Identifications with $K\simlt$19
were observed with the 3.6\,m ESO New Technology Telescope (NTT), and the faintest targets ($K>19$) were observed
using the 8\,m ESO Very Large Telescope (VLT). For both the NTT and VLT observations,
to acquire the faint targets into the slit, we
used blind offsets from a nearby ($\simlt$1\,arcmin) star in the
$K$-band images. 
After a first exposure, we performed a
quick data reduction. If the redshift was obvious, the next exposure
was aborted to save observing time. 
The individual exposures were
offset by 10\,arcsec along the slit to allow subtraction of the
fringing in the detector at red wavelengths.

The journal of observations is given in Table~\ref{obslog}.

\subsubsection{ANU 2.3m}
\label{sec:2.3m}

We used the Dual-Beam Spectrograph  \citep[DBS; ][]{rcb88} on the
ANU's 2.3-m telescope on 2007 April 10--14. A plane mirror was used to
direct all the light into the blue arm of the spectrograph in order to
obtain complete wavelength coverage over the range 3500--11000~\AA.
The CCD is an E2V $2148 \times 562$ pixel detector with a spatial
scale of 0.91\,arcsec\,pixel$^{-1}$. We used the 158R grating, providing a
dispersion of 4\,\AA\,pixel$^{-1}$ and a spectral resolution of
8.9\,\AA. Severe fringing at the red end restricted the usable
wavelength range to 3500--8000\,\AA; second order contamination was minimal. Conditions were variable, but
mostly non-photometric, with poor seeing.

\subsubsection{ESO/NTT}
\label{sec:NTT}

We used the ESO Multi-Mode Instrument \citep[EMMI;][]{Dek86} on the
NTT on 2006 July 25--27 and 2007 July 10--12.
The conditions in 2006 were poor with the first two nights lost to weather;
the third was non-photometric with 1.0--1.4-arcsec
seeing. In 2007 the conditions were better, with seeing between 0.6
and 1.0-arcsec FWHM. We used grism \#2 and a 1.5- or 1-arcsec slit; the latter
was used in the better seeing conditions. The wavelength coverage was
3700--9700~\AA, with a 
dispersion of 3.5 \AA\ pixel$^{-1}$, a spectral resolution of 10.4~\AA\
FWHM and the pixel size was 0.33\,arcsec. To minimise the effects of
differential atmospheric refraction, we observed the targets with the slit 
oriented at the
parallactic position angle.

\subsubsection{ESO/VLT}
\label{sec:VLT}

We used the FOcal Reducer and Spectrograph \citep[FORS; ][]{aff+98}
in visitor mode on the VLT on 2006 June 22--23. 
We used FORS2 on UT1 (Antu) with the 150I grism and 1-arcsec
slit, providing a dispersion of 6.7\,\AA\,pixel$^{-1}$ and a spectral
resolution of 21.4\,\AA. The wavelength coverage was 3600--8500\,\AA.
As the FORS instruments have a linear atmospheric
dispersion corrector, we did not observe at the parallactic angle, and 
instead the slit position angle was chosen to avoid bright stars. 

\begin{table}
\caption{Log of observations.}
\label{obslog}
\begin{center}
\begin{tabular}{lccc}
\hline
 Date & Telescope & Instrument  & Seeing\\
  &           &            &     (arcsec)   \\
\hline
 2006 June 22--23 & VLT & FORS2  & 0.5--1.3 \\
 2006 July 27 & NTT & EMMI & 1.0--1.4 \\
 2007 April 10--14 & ANU 2.3\,m & DBS  & 1.5--3.0  \\
 2007 July 10--12 & NTT & EMMI  & 0.6--1.0 \\
\hline
\end{tabular}
\end{center}
\end{table}

\subsubsection{Data reduction}
\label{specred}

Data reduction was performed with the \textsc{iraf} software suite,
using standard procedures. We removed the bias and pixel-to-pixel gain
variations from each frame, and then removed the cosmic rays using the
\textsc{iraf} task \textsc{szap}. We extracted spectra using the
\textsc{iraf} task \textsc{apall}, extracting only the central few
rows ($\sim 1$\,arcsec) to maximise the signal-to-noise ratio of the
spectrum. This optimises our chance of finding a redshift, at the cost
of not including all the flux from the object, particularly from any
extended line emission. The same extraction aperture was used to extract a
calibration spectrum from the arc lamp, which was then used to
calibrate the one-dimensional spectrum in wavelength. 

The spectra were flux-calibrated by comparing with the spectrum of a
spectrophotometric standard taken on the same night. 
No correction was made for Galactic reddening.

\section{RESULTS}
\label{results}

Table~\ref{data} lists all the MRCR--SUMSS sources for which we
have taken spectra. The $K$-band images with radio contours for
these objects were presented in \citetalias{Bry08}. 
Some sources met our selection 
criteria, but were also part of the SUMSS--NVSS sample \citep[][Paper\,I]{deB04},
and were therefore not reobserved. For those sources, the $K$-band magnitudes and 
positions given are from \citet{deB04}. 

The columns in Table~\ref{data} 
are:

\noindent (1) NVSS source name. 

\noindent (2) $K$-band magnitude in a 4-arcsec-diameter aperture. Some sources are obscured by foreground
stars (see column\,6) which contaminate the listed magnitude.
Sources marked SC were visible on the SuperCOSMOS UKR fields and no $K$-band image was obtained 
\citepalias[see][]{Bry08}.
Our analysis exclusively used magnitudes in a 4-arcsec aperture,
and the justification for this choice is given in \citetalias{Bry08}.

\noindent (3) $K$-band magnitudes in equivalent 64-kpc apertures, which were
calculated from 8-arcsec-aperture magnitudes, following the method of 
\citet{eal97}. 
Only radio galaxies (not QSOs) which have redshifts and 
8-arcsec-aperture magnitudes in
\citetalias{Bry08} have been converted. The limitations on these 64-kpc-aperture
magnitudes were discussed in \citetalias{Bry08} and are discussed further in Section~\ref{Kztrends}.

\noindent (4) \& (5) RA and DEC (J2000) of the $K$-band identification. 
One source which had no $K$-band identification is marked ``---".
The sources which had SuperCOSMOS identifications have the SuperCOSMOS
position listed.

\noindent (6) Spectroscopic redshift. 
The wavelength of each line was determined by fitting a Gaussian on top of a continuum. 
Occasionally spurious lines were introduced by the fringing in the CCD; we required that
features be visible in both exposures to be counted as real.
The redshift quoted is the average of the redshifts derived from individual lines; the
error is the scatter in this average, with uncertainties due to the line fitting
and wavelength calibration added in quadrature.
The sources marked `continuum' 
had continuum but no lines, while those marked `undetected' showed no
continuum or line emission. Some objects are obscured by a foreground star and are listed as
`obscured'. Objects with emission lines broader than $>3000$\,km\,s$^{-1}$ have been classified as 
QSOs
\footnote{NVSS~J144932$-$385657 has one line with FWHM$>3000$\,km\,s$^{-1}$ 
but it is faint
and the error on the width is large; since the other lines are narrow we do not classify it as a QSO.} 
and are marked with a `Q'.
 
\noindent (7) Origin of the spectra. On the NTT, EMMI was the
spectrograph, while FORS2 was the instrument on the VLT. `DB' refers to objects
from \citet{deB04,deB06}.

\noindent (8) Linear size in kpc using the largest angular size measurements
at 2368\,MHz unless marked with a footnote.

\noindent (9) Rest-frame luminosity at 1.4\,GHz in W\,Hz$^{-1}$. 
Luminosities were calculated using the 5-point spectral index
from \citetalias{Bry08} (table\,2) unless marked by a footnote.
 
  
Of the 164 sources with $K$-band or SuperCOSMOS counterparts in \citetalias{Bry08},
spectra were obtained for 52 objects listed in Table~\ref{data}, including 46 new spectra and
six that are common to the SUMSS--NVSS sample and therefore have spectra
in \citet{deB06}.
NVSS~J231727$-$352606 is in both the SUMSS--NVSS and MRCR--SUMSS samples, 
and has a $K$-band magnitude 
limit listed in \citet{deB04} and a redshift from \citet{deB06}; our new, deeper limiting magnitude 
\citepalias[see][]{Bry08} is
shown in the table. NVSS~J103441$-$394957 and the combined source 
NVSS~J023601$-$314204/J023605$-$314235 have very faint red identifications in
SuperCOSMOS, and were retained in the sample
because the identification was not located until we had high-resolution ATCA images.

The spectroscopy has resulted in 36 confirmed redshifts, 12 continuum spectra, one that remained
undetected and
three that are obscured by stars.
The spectra are shown in Fig.~\ref{fig:spectra}; they were smoothed lightly for the plots, 
using (typically) a 3-pixel boxcar.

\begin{table*}
\begin{minipage}{160mm}
\caption{Spectroscopy results including redshifts and the derived linear sizes and rest-frame luminosities. Details of each column are given in Section 3.}
\label{data}
\scriptsize
\begin{tabular}{lcccccccc}\hline
\hline

(1)       &              (2)  &  (3)   &          (4)    &  (5)     &      (6)  &          (7)            &       (8)  & (9)              \\
Name &  $K$    &  $K$ & RA$_{{\rm J2000}}$ & DEC$_{{\rm J2000}}$ & Redshift & Origin   & Radio  & L$_{1.4\,{\rm GHz}}$  \\
 &  (4$''$ ap.)   &  (64\,kpc) & $K$-band & $K$-band &  &   &  linear &    \\
  &  (mag) & (mag)  & (h m s)  & ($^{\circ}$ $'$  $''$) &     &  & size (kpc) & (W\,Hz$^{-1}$)  \\
\hline
NVSS J001506$-$330155 &  $ 17.6 \pm 0.2 $ & & 00:15:06.31 & $-$33:01:54.0 &  continuum &  NTT &  &   \\
NVSS J003445$-$372348 &  $ 17.0 \pm 0.1 $ & & 00:34:45.66 & $-$37:23:47.4 &  continuum &  NTT &  &   \\
NVSS J011606$-$331241$^{\it i}$ &  $ 18.6 \pm 0.2 $ & 18.0 &  01:16:06.77 & $-$33:12:42.8 & $ 0.352 \pm 0.001 $ &   DB & 31.5 & $9.9\times10^{24}$  \\
NVSS J023601$-$314204\vline$^{\it a}$ &    SC & & 02:36:03.99 & $-$31:42:22.2\rlap{$^{\it b}$} & $ 0.500 \pm 0.001 $ &  NTT & 357.7 & $5.1\times10^{25}$  \\
NVSS J023605$-$314235\vline$^{\it a}$ & &       &  &   &   &       &      &   \\
NVSS J101008$-$383629 &  $ 18.9 \pm 0.2 $ &  & 10:10:08.02 & $-$38:36:29.2 & continuum &  VLT &  &   \\
NVSS J103441$-$394957 &   SC     &  & 10:34:42.10 & $-$39:49:58.2\rlap{$^{\it b}$}  & $ 1.832 \pm 0.008$ {\bf Q}   &  2.3\,m &  247.3 & $1.3\times10^{27}$   \\
NVSS J103615$-$321659 &  $ 19.3 \pm 0.2 $ & &  10:36:15.26 & $-$32:16:57.4 & $ 2.136 \pm 0.012 $ &  VLT &  87.5 & $3.0\times10^{27}$\rlap{$^{\it n}$}  \\
NVSS J105917$-$303658 &  $ 18.8 \pm 0.1 $ & & 10:59:17.45 & $-$30:36:57.5 & $ 3.263 \pm 0.022 $ {\bf Q} &  VLT & 7.6\rlap{$^{\it n}$} & $8.1\times10^{27}$\rlap{$^{\it n}$}  \\
NVSS J111921$-$363139 &  $ 18.7 \pm 0.1 $ & 19.1 & 11:19:21.81 & $-$36:31:39.1 & $ 2.768 \pm 0.006 $ &  VLT &  43.2\rlap{$^{\it n}$} & $6.1\times10^{28}$\rlap{$^{\it n}$}  \\
\vspace{3mm}
NVSS J112641$-$383950 &  $ 20.3 \pm 0.3 $ & & 11:26:41.00 & $-$38:39:53.5 & $ 1.856 \pm 0.003 $ &  VLT &  9.4\rlap{$^{\it n}$} & $3.2\times10^{27}$\rlap{$^{\it n}$}  \\
NVSS J120839$-$340307 &  $ 18.0 \pm 0.1 $ & 17.7 & 12:08:39.75 & $-$34:03:09.6 & $ 1.120 \pm 0.002 $ &  NTT &  111.3 & $9.0\times10^{27}$  \\
NVSS J122553$-$382823 &  $ 19.2 \pm 0.2 $ & & 12:25:53.43 & $-$38:28:26.2 &  continuum  &  VLT &   &   \\
NVSS J140022$-$344411 &  $ 16.3 \pm 0.2 $ & 16.1 & 14:00:23.01 & $-$34:44:11.3 & $ 0.7844 \pm 0.0007 $ &  NTT &  64.1 & $2.5\times10^{26}$  \\
NVSS J140223$-$363539 &  $ 19.8 \pm 0.3 $ & 19.9 & 14:02:23.63 & $-$36:35:42.2 & $ 2.796 \pm 0.012 $ &  VLT &  25.6\rlap{$^{\it n}$} & $5.0\times10^{27}$\rlap{$^{\it n}$}  \\
NVSS J140854$-$382731 &  $ 16.9 \pm 0.2 $ & 16.6 & 14:08:55.24 & $-$38:27:38.7 & $ 0.870 \pm 0.003 $ & NTT & 43.3\rlap{$^{\it n}$} & $4.2\times10^{26}$\rlap{$^{\it n}$}  \\
NVSS J141428$-$320637 &  $ 15.5 \pm 0.2 $\rlap{$^{\it e}$} & 15.4& 14:14:28.27 & $-$32:06:38.7\rlap{$^{\it c}$}  & $ 0.2775 \pm 0.0008 $ &  2.3\,m & 6.7 & $4.4\times10^{25}$  \\
NVSS J144206$-$345115 & $>$   20.1    &  & 14:42:06.69 & $-$34:51:15.6 & $ 2.750 \pm 0.004 $ {\bf Q} &  VLT &  7.2\rlap{$^{\it n}$} & $2.9\times10^{27}$\rlap{$^{\it n}$}  \\
NVSS J144932$-$385657 &  $ 19.8 \pm 0.2 $ & & 14:49:32.79 & $-$38:56:57.5 & $ 2.149 \pm 0.002 $ &  VLT &  57.2\rlap{$^{\it n}$} & $3.4\times10^{27}$\rlap{$^{\it n}$}  \\
NVSS J145913$-$380118 &  $ 18.3 \pm 0.1 $ & & 14:59:12.99 & $-$38:01:19.0 &  continuum &  NTT &   &   \\
\vspace{3mm}
NVSS J151020$-$352803 &  $ 20.2 \pm 0.2 $ & 20.5& 15:10:20.84 & $-$35:28:03.2 & $ 2.931 \pm 0.006 $ &  VLT &  28.4\rlap{$^{\it n}$} & $5.2\times10^{27}$\rlap{$^{\it n}$}  \\
NVSS J151021$-$364253 &  $ 20.1 \pm 0.2 $ & & 15:10:21.80 & $-$36:42:54.1 &  continuum &  VLT &   &   \\
NVSS J151215$-$382220 &  $ 17.2 \pm 0.1 $ & & 15:12:15.51 & $-$38:22:21.7 &  continuum &  2.3\,m &  &   \\
NVSS J151503$-$373511 &  $ 17.3 \pm 0.1 $ & & 15:15:03.09 & $-$37:35:12.2 &  continuum & NTT &   &   \\
NVSS J152123$-$375708 &  $ 19.3 \pm 0.3 $ & & 15:21:24.02 & $-$37:57:10.2 & $ 1.530 \pm 0.003 $ {\bf Q} &  VLT &  31.0 & $6.3\times10^{26}$  \\
NVSS J152435$-$352623 &  $ 20.9 \pm 0.5 $ &  &15:24:35.41 & $-$35:26:22.0 & $ 2.752 \pm 0.007 $ {\bf Q} &  VLT &  16.8\rlap{$^{\it n}$} & $6.4\times10^{27}$\rlap{$^{\it n}$}   \\
NVSS J152445$-$373200 &  $ 18.3 \pm 0.1 $ & 18.6 & 15:24:45.37 & $-$37:31:58.5 & $ 1.798 \pm 0.008 $ & NTT &  95.6\rlap{$^{\it n}$} & $1.4\times10^{27}$\rlap{$^{\it n}$}  \\
NVSS J152737$-$301459 &  $ 17.6 \pm 0.2 $ & & 15:27:37.88 & $-$30:14:58.6 & $ 0.969 \pm 0.008 $ & VLT &  105.8 & $1.0\times10^{27}$  \\
NVSS J152747$-$364218 &  $ 19.6 \pm 0.2 $ & & 15:27:47.38 & $-$36:42:18.4 & obscured by star &  VLT &   &   \\
NVSS J152921$-$362209 &  $ 15.7 \pm 0.3$\rlap{$^{\it e}$} & & 15:29:21.77 & $-$36:22:14.6\rlap{$^{\it c}$} & obscured by star &   2.3\,m &   &   \\
\vspace{3mm}
NVSS J202856$-$353709$^{\it i}$ &  $ 16.9 \pm 0.1 $ & & 20:28:56.77 & $-$35:37:06.0 &  obscured by star  &   DB &   &   \\
NVSS J202945$-$344812$^{\it i}$ &  $ 17.6 \pm 0.1 $ & 17.3 & 20:29:45.82 & $-$34:48:15.5 & $ 1.497 \pm 0.002 $ &  DB &   161.4\rlap{$^{\it m}$} & $7.2\times10^{26}$\rlap{$^{\it m}$}  \\
NVSS J210814$-$350823 &  $ 19.5 \pm 0.5 $ & &  21:08:14.46 & $-$35:08:25.3 & $ 1.879 \pm 0.002 $ &  VLT &  50.3\rlap{$^{\it n}$} & $3.3\times10^{27}$\rlap{$^{\it n}$}  \\
NVSS J212048$-$333214 &  $ 19.1 \pm 0.1 $ & & 21:20:48.76 & $-$33:32:14.4 &  continuum  &  NTT &   &   \\
NVSS J213238$-$335318 &  $ 19.8 \pm 0.4 $ & & 21:32:38.95 & $-$33:53:18.9 & $ 2.900 \pm 0.001 $ &  VLT &  13.4\rlap{$^{\it n}$} & $9.4\times10^{27}$\rlap{$^{\it n}$}  \\
NVSS J213434$-$302522 &  $ 16.8 \pm 0.2 $ & 16.7 & 21:34:34.20 & $-$30:25:20.6 & $ 0.6819 \pm 0.0005 $ &  NTT &  57.5 & $1.2\times10^{26}$  \\
NVSS J213637$-$340318 &  $ 19.7 \pm 0.2 $ & 19.6 & 21:36:37.39 & $-$34:03:21.1 & $ 2.772 \pm 0.002 $ &  VLT &  25.6\rlap{$^{\it n}$} & $7.2\times10^{27}$\rlap{$^{\it n}$}  \\
NVSS J215009$-$341052 &  $ 19.6 \pm 0.3 $ & & 21:50:09.30 & $-$34:10:52.4 & continuum  &  VLT &  &  \\
NVSS J215047$-$343616 &  $ 17.4 \pm 0.1 $ & & 21:50:47.38 & $-$34:36:17.5 &  continuum  &  NTT  &   &   \\
NVSS J215226$-$341606 &  $ 18.3 \pm 0.3 $ & 18.0 & 21:52:26.74 & $-$34:16:06.2 & $ 1.277 \pm 0.002 $ & NTT &  180.0 & $7.7\times10^{26}$  \\
\vspace{3mm}
NVSS J215455$-$363006 &  $ 17.7 \pm 0.1 $ & 16.9 & 21:54:55.08 & $-$36:30:06.8 & $ 1.235 \pm 0.001 $ & NTT &  43.7 & $3.4\times10^{27}$  \\
NVSS J221104$-$351829 &  $ 19.0 \pm 0.2 $ &  & 22:11:05.04 & $-$35:18:28.8 &  continuum &  VLT &   &   \\
NVSS J223101$-$353227 &  $ 19.5 \pm 0.2 $ & & 22:31:01.72 & $-$35:32:28.8 & $ 1.834 \pm 0.006 $ {\bf Q} & VLT &  23.2 & $2.3\times10^{27}$   \\
NVSS J223111$-$371459 &  $ 17.6 \pm 0.1 $ & 17.3 & 22:31:11.09 & $-$37:14:59.9 & $ 1.248 \pm 0.002 $ & NTT &  20.4 & $7.2\times10^{26}$  \\
NVSS J223305$-$365658 &  $ 16.6 \pm 0.1 $ & 16.3 & 22:33:05.08 & $-$36:56:58.2 & $ 0.939 \pm 0.002 $ & NTT &  35.1 & $7.5\times10^{26}$  \\
NVSS J225719$-$343954$^{\it i}$ &  $ 16.7 \pm 0.1 $ & 16.5 & 22:57:19.63 & $-$34:39:54.6 & $ 0.726 \pm 0.001 $ &   DB &   $<$43.5\rlap{$^{\it m}$} & $1.4\times10^{26}$\rlap{$^{\it m}$}  \\
NVSS J230004$-$304711 &  $ 17.4 \pm 0.1 $ & 16.9 & 23:00:04.31 & $-$30:47:08.7 & $ 0.5445 \pm 0.0006 $ &  NTT &  97.5 & $7.6\times10^{25}$  \\
NVSS J231727$-$352606$^{\it h}$ & $>$   20.7    &  &  --- & --- & $ 3.874 \pm 0.002 $ & DB &  28.7\rlap{$^{\it m}$} & $1.4\times10^{28}$  \\
NVSS J232125$-$375829 &  $ 17.9 \pm 0.1 $ & & 23:21:25.55 & $-$37:58:30.7 & $ 2.204 \pm 0.006 $ {\bf Q} &  NTT &  $<$8.4\rlap{$^{\it n}$} & $4.0\times10^{27}$\rlap{$^{\it n}$}  \\
NVSS J233034$-$330009 &  $ 17.2 \pm 0.3 $\rlap{$^{\it l}$} & 17.1 & 23:30:34.49 & $-$33:00:11.5 & $ 2.675 \pm 0.005 $ &  NTT &  123.5\rlap{$^{\it n}$} & $6.4\times10^{27}$\rlap{$^{\it n}$}  \\
\vspace{3mm}
NVSS J233226$-$363423 &  $ 17.9 \pm 0.1 $ &  & 23:32:25.96 & $-$36:34:24.1 & $ 0.988 \pm 0.003 $ {\bf Q} & NTT &  85.1 & $3.3\times10^{26}$  \\
NVSS J233627$-$324323 &  $ 18.5 \pm 0.1 $ & & 23:36:27.09 & $-$32:43:21.8 &  undetected &  NTT &   &   \\
NVSS J234145$-$350624$^{\it i}$  &  $ 16.9 \pm 0.1 $ & 16.2 & 23:41:45.85 & $-$35:06:22.2 & $ 0.644 \pm 0.001 $ &  DB &   $<$34.5\rlap{$^{\it m}$} & $3.4\times10^{27}$\rlap{$^{\it m}$}  \\
\hline
\end{tabular}
\vspace{-0.7cm}
\footnotetext[1]{Two NVSS sources have been shown to be the components of one radio source.}
\footnotetext[2]{SuperCOSMOS UKR optical position. \hspace{0.5cm}$^{\it c}$ 2MASS \citep{Skr06} position.}
\footnotetext[5] {2MASS $K$ magnitude in 8-arcsec aperture.} 
\footnotetext[8] {This source was observed by \citet{deB04} in the SUMSS--NVSS sample with a $K$-band non-detection. We have reobserved it and improved the limit on the $K$-band non-detection.}
\footnotetext[9] {These sources were also part of the SUMSS--NVSS sample. The $K$-band magnitudes and positions 
given are from \citet{deB04}.}
\footnotetext[12] {This object is at the position where the redshift was measured, but it may be a foreground star (see Section~\ref{notes}).}
\footnotetext[13] {Spectral indices used in calculating the luminosities are from \citet{kla06} 
and the linear sizes are derived from LAS measurements either in 
\citet{deB04} or \citet{kla06}.
}
\footnotetext[14] {Linear size is based on the largest angular size at 4800 or 8640\,MHz. Luminosity was calculated using the 7-point spectral index (including 4800- and 8640-MHz data) from 
\citetalias{Bry08} (table\,3).}
\end{minipage}
\end{table*}

Parameters of the emission lines were measured following 
procedures similar to those described in \citet{Rott97}; these parameters are listed in
Table~\ref{tab:lineparams}. The line centre and width (FWHM) were
determined by fitting a Gaussian on top of a continuum; the line flux
was obtained by direct summation of the flux above the continuum over
a wavelength range four times the FWHM, except where such a width
would also include neighbouring lines. No attempt has been made to
de-blend lines. The location of the continuum was determined by eye
separately for each line. This is the major source of error in the
line ratios. The errors on the line fluxes and widths were estimated by 
comparing the measured values using different continuum levels; for some of 
the weaker lines, these errors are substantial. 
Because we extracted the spectra from only the central
few rows of each object (see Section~\ref{specred}) we have
not in general included all the flux from each emission line, so the line ratios we
derive are characteristic of the nuclear regions only.

\begin{table*}
\caption{Emission line measurements.}
\label{tab:lineparams}
\begin{tabular}{l r@{$\;\pm\;$}l l r@{$\;\pm\;$}l r@{$\;\pm\;$}l
r@{$\;\pm\;$}l}
\hline
\hline
       & \multicolumn{2}{c}{ }   &      &
\multicolumn{2}{c}{$\lambda_\mathrm{obs}$} & \multicolumn{2}{c}{$\Delta
v_\mathrm{FWHM}$} & \multicolumn{2}{c}{Flux} \\
Source & \multicolumn{2}{c}{$z$} & Line & \multicolumn{2}{c}{(\AA)}
& \multicolumn{2}{c}{(km s$^{-1}$)}            & \multicolumn{2}{c}{($\times
10^{-16}\;\mathrm{erg}\;\mathrm{s}^{-1}\;\mathrm{cm}^{-2}$)} \\
\hline
NVSS J103441$-$394957 & 1.832 & 0.008         & \ion{C}{iv}\   $\lambda1549$ & 4389.0 & 1.2 &  6350 & 700     
            & 162   & 40 \\
                      & \multicolumn{2}{c}{ } & \ion{C}{iii}]\ $\lambda1909$ & 5368.4 & 9.7 & 14900 &5000     
            &  99   & 25 \\
NVSS J103615$-$321659 & 2.136 & 0.012         & Ly$\;\alpha$\ $\lambda1216 $ & 3803.0 & 0.3 & \multicolumn{2}{c}{$<1700$} & 0.85 & 0.21 \\
                      & \multicolumn{2}{c}{ } & \ion{C}{iv}\ $\lambda1549  $ & 4871.3 & 0.5 &  1580 & 160     & 0.30 & 0.08 \\
NVSS J105917$-$303658 & 3.263 & 0.022         & Ly$\;\beta$\ $\lambda1026  $ & 4341.1 & 3.4 &  7500 & 1000     
            & 1.6  & 0.4  \\
                      & \multicolumn{2}{c}{ } & Ly$\;\alpha$\ $\lambda1216 $ & 5219.8 & 1.0 &  4200 & 300     
            & 1.9  & 0.5 \\
                      & \multicolumn{2}{c}{ } & \ion{C}{iv}\ $\lambda1549  $ & 6611.4 & 2.6 &  7300 & 600     
            & 2.6  & 0.6 \\
                      & \multicolumn{2}{c}{ } & \ion{He}{ii}\ $\lambda1640 $ & 6991.2 & 0.4 &  1300 & 80      
            & 1.3  & 0.3 \\
                      & \multicolumn{2}{c}{ } & \ion{C}{iii}]\ $\lambda1909$ & 8124.8 & 0.9 &  1600 & 170      
            & 1.5  & 0.4 \\
NVSS J111921$-$363139 & 2.768 & 0.006         & Ly$\;\alpha$\ $\lambda1216 $ & 4588.6 & 0.1 &  270 & 20       
            & 1.8  & 0.5 \\
                      & \multicolumn{2}{c}{ } &\ion{C}{iv}\ $\lambda1549   $ & 5841.1 & 0.7 & 2700 & 180       
            & 1.4  & 0.3 \\
                      & \multicolumn{2}{c}{ } &\ion{He}{ii}\ $\lambda1640  $ & 6173.8 & 0.4 & 1450 & 90       
            & 0.9  & 0.2 \\
                      & \multicolumn{2}{c}{ } &\ion{C}{iii}]\ $\lambda1909 $ & 7179.3 & 0.6 & 1200 & 130       
            & 0.9  & 0.2 \\
NVSS J112641$-$383950 & 1.856 & 0.003         & \ion{C}{iv}   $\lambda1549 $ & 4429.8 & 0.1 &  750 &  40      
            & 0.6  & 0.2  \\
                      & \multicolumn{2}{c}{ } & \ion{He}{ii}  $\lambda1640 $ & 4687.4 & 0.2 &  480 &  60      
            & 0.4  & 0.1   \\
                      & \multicolumn{2}{c}{ } & \ion{Al}{iii}] $\lambda1858$ & 5299.7 & 1.3 & 1100 &  400     
            & 0.05 & 0.03  \\
                      & \multicolumn{2}{c}{ } & \ion{C}{iii}] $\lambda1909 $ & 5448.4 & 0.1 &  810 &  30      
            & 0.4  & 0.1  \\
                      & \multicolumn{2}{c}{ } & \ion{C}{ii}] $\lambda2326 $  & 6647.2 & 0.4 & 1270 &  90      
            & 0.4  & 0.2 \\
                      & \multicolumn{2}{c}{ } & [\ion{Ne}{iv}] $\lambda2423$ & 6921.2 & 0.5 & \multicolumn{2}{
c}{$<930$}  & 0.15 & 0.04 \\
                      & \multicolumn{2}{c}{ } & \ion{Mg}{ii} $\lambda2798 $  & 7998.0 & 0.3 & 1120 &  70      
            & 0.9  & 0.2  \\
NVSS J120839$-$340307 & 1.120 & 0.002         & \ion{Mg}{ii} $\lambda2798 $  & 5932.7 & 0.8 & 1750 & 450      
             & 5.2  & 3 \\
                      & \multicolumn{2}{c}{ } & [\ion{O}{ii}]  $\lambda3727$ & 7906.7 & 0.4 &  800 & 100      
             & 7.5  & 4 \\
NVSS J140022$-$344411 & 0.7844 & 0.0007         & [\ion{O}{ii}] $\lambda3727 $ & 6652.4 & 0.3 & \multicolumn{2
}{c}{$<470$}  & 7.6  & 4 \\
NVSS J140223$-$363539 & 2.796 & 0.012         & Ly$\;\alpha$\ $\lambda1216 $ & 4603.6 & 0.3 &  1630 &  100     
            & 0.9  & 0.2 \\
                      & \multicolumn{2}{c}{ } & \ion{C}{iv}  $\lambda1549 $  & 5892.1 & 0.3 & \multicolumn{2}{
c}{$<1100$} & 0.06 & 0.03 \\
NVSS J140854$-$382731 & 0.870 & 0.003         & [\ion{O}{ii}]  $\lambda3727$ & 6966.4 & 0.2 &  470 &  90      
            & 0.79  & 0.4 \\
                      & \multicolumn{2}{c}{ } & [\ion{Ne}{iii}] $\lambda3869$& 7214.3 & 1.9 & 1580 & 700      
            & 0.57  & 0.3 \\
NVSS J141428$-$320637 & 0.2775 & 0.0008         & [\ion{O}{ii}]  $\lambda3727$ & 4759.8 & 0.1 &  660 &  50    
              & 125 & 30 \\
                      & \multicolumn{2}{c}{ } & [\ion{O}{iii}] $\lambda4363$ & 5576.2 & 0.1 & \multicolumn{2}{
c}{$<480$} & 5.7 & 3 \\
                      & \multicolumn{2}{c}{ } & H$\beta$ $\lambda4861      $ & 6209.9 & 0.1 &  110 &  50      
            &  16.9 & 4 \\
                      & \multicolumn{2}{c}{ } & [\ion{O}{iii}] $\lambda4959$ & 6331.6 & 0.4 &  560 &  200     
             &  13.2 & 3 \\
                      & \multicolumn{2}{c}{ } & [\ion{O}{iii}] $\lambda5007$ & 6393.5 & 0.1 &  470 &  50      
            &  49.9 & 10 \\
                      & \multicolumn{2}{c}{ } & [\ion{N}{ii}] $\lambda6583$ + H$\alpha$ $\lambda6562$ & 8392.5
 & 0.5 & 2500 &  100  & 180 & 45 \\
NVSS J144206$-$345115 & 2.750 & 0.004         & Ly$\;\beta$\ $\lambda1025$ + \ion{O}{vi} $\lambda 1034$ & 3896.2 & 2.3 & 8200 & 900 &  2.6 & 0.6  \\
                      & \multicolumn{2}{c}{ } & Ly$\;\alpha$\ $\lambda1216 $ & 4563.8 & 0.1 & 2840 & 50  &  6.1 & 1.5   \\
    & \multicolumn{2}{c}{ } & \ion{Si}{iv} 1400 + \ion{O}{iv}] $\lambda1402$ & 5252.5 & 2.3 & 6100 & 700 &  1.1 & 0.3  \\
                      & \multicolumn{2}{c}{ } & \ion{C}{iv}   $\lambda1549 $ & 5808.1 & 1.2 & 6800 & 300 &  2.8 & 0.7   \\
                      & \multicolumn{2}{c}{ } & \ion{He}{ii}  $\lambda1640 $ & 6153.0 & 9.2 & 5800 & 2000 &  0.3 & 0.1 \\
                      & \multicolumn{2}{c}{ } & \ion{C}{iii}] $\lambda1909 $ & 7142.3 & 1.0 & 5600 & 200  &  1.8 & 0.4  \\
NVSS J144932$-$385657 & 2.149 & 0.002         & Ly$\;\alpha$\ $\lambda1216 $ & 3832.1 & 0.2 & 2730 & 70  & 1.9 & 0.5  \\
                      & \multicolumn{2}{c}{ } & \ion{C}{iv}  $\lambda1549  $ & 4876.0 & 0.2 & 390  & 50  & 0.4 & 0.1   \\
                      & \multicolumn{2}{c}{ } & \ion{He}{ii} $\lambda1640  $ & 5164.6 & 6.4 & 4700 & 2000 & 0.26 & 0.06 \\
NVSS J151020$-$352803 & 2.931 & 0.006         & Ly$\;\alpha$\ $\lambda1216 $ & 4782.2 & 0.1 & 1340 & 40  & 1.4 & 0.3 \\
                      & \multicolumn{2}{c}{ } & \ion{C}{iv}   $\lambda1549 $ & 6097.8 & 2.2 & 1700 & 500 & 0.11 & 0.03 \\
                      & \multicolumn{2}{c}{ } & \ion{He}{ii}  $\lambda1640 $ & 6436.9 & 0.9 & 1330 & 200  & 0.11 & 0.03 \\
                      & \multicolumn{2}{c}{ } & \ion{C}{iii}] $\lambda1909 $ & 7503.5 & 0.5 & 1110 & 100  & 0.25 & 0.06 \\
NVSS J152123$-$375708 & 1.530 & 0.003         & \ion{C}{iv}   $\lambda1549 $ & 3913.6 & 1.2 & 8900 & 400 & 3.5 & 0.9 \\
                      & \multicolumn{2}{c}{ } & \ion{C}{iii}] $\lambda1909 $ & 4831.5 & 0.6 & 2600 & 200  & 0.6 & 0.2 \\
                      & \multicolumn{2}{c}{ } & \ion{Mg}{ii}  $\lambda2798 $ & 7084.8 & 2.5 & 5700 & 500 & 1.3 & 0.3 \\
NVSS J152435$-$352623 & 2.752 & 0.007         & Ly$\;\alpha$\ $\lambda1216 $ & 4571.5 & 0.3 & 4430 & 100  & 3.1 & 0.8 \\
                      & \multicolumn{2}{c}{ } & \ion{C}{iv}   $\lambda1549 $ & 5803.0 & 6.1 &15800 & 1500 & 2.5 & 0.6 \\
                      & \multicolumn{2}{c}{ } & \ion{C}{iii}] $\lambda1909 $ & 7157.7 & 3.8 & 7700 & 800 & 1.8 & 0.4 \\
NVSS J152445$-$373200 & 1.798 & 0.008         & \ion{C}{iv} $\lambda1549   $ & 4334.7 & 0.6 & 1530 &  400 &  9.5  & 5 \\
                      & \multicolumn{2}{c}{ } & [\ion{Ne}{v}] $\lambda3345 $ & 9363.4 & 0.4 &  740 &  100 & 20.8  & 10 \\
NVSS J152737$-$301459 & 0.969 & 0.008         & [\ion{O}{ii}] $\lambda3727 $ & 7341.0 & 0.3 &  340 &  70 & 0.61  & 0.15 \\

\hline
\end{tabular}
\end{table*}

\begin{table*}
\contcaption{}
\begin{tabular}{l r@{$\;\pm\;$}l l r@{$\;\pm\;$}l r@{$\;\pm\;$}l
r@{$\;\pm\;$}l}
\hline
\hline
       & \multicolumn{2}{c}{ }   &      &
\multicolumn{2}{c}{$\lambda_\mathrm{obs}$} & \multicolumn{2}{c}{$\Delta
v_\mathrm{FWHM}$} & \multicolumn{2}{c}{Flux} \\
Source & \multicolumn{2}{c}{$z$} & Line & \multicolumn{2}{c}{(\AA)}
& \multicolumn{2}{c}{(km s$^{-1}$)}            & \multicolumn{2}{c}{($\times
10^{-16}\;\mathrm{erg}\;\mathrm{s}^{-1}\;\mathrm{cm}^{-2}$)} \\
\hline
NVSS J210814$-$350823 & 1.879 & 0.002         & \ion{C}{iv} $\lambda1549   $ & 4462.1 & 0.3 & 1730  &  100     & 0.9  & 0.2 \\
                      & \multicolumn{2}{c}{ } & \ion{He}{ii} $\lambda1640  $ & 4721.2 & 0.2 &  830  &  70     & 0.7  & 0.2 \\
                      & \multicolumn{2}{c}{ } & \ion{C}{iii}] $\lambda1909 $ & 5491.1 & 0.1 & \multicolumn{2}{c}{$<1200$}  & 0.5  & 0.1 \\
                      & \multicolumn{2}{c}{ } & \ion{C}{ii}] $\lambda2326  $ & 6694.1 & 0.6 & 1340  &  130     & 0.24 & 0.06 \\
                      & \multicolumn{2}{c}{ } & [\ion{Ne}{iv}] $\lambda2423$ & 6977.5 & 0.4 &  900  &  80     & 0.4  & 0.1 \\
                      & \multicolumn{2}{c}{ } & \ion{Mg}{ii} $\lambda2798  $ & 8060.1 & 0.4 & \multicolumn{2}{c}{$<800$}   & 0.4  & 0.2 \\
NVSS J213238$-$335318 & 2.900 & 0.001         & Ly$\;\alpha$\ $\lambda1216 $ & 4740.6 & 0.5 & 2100 &  200      & 0.97 & 0.2 \\
NVSS J213434$-$302522 & 0.6819 & 0.0005         & [\ion{O}{ii}] $\lambda3727$  & 6273.1 & 0.2 &  310 &  90    & 3.6   & 2 \\
NVSS J213637$-$340318 & 2.772 & 0.002         & Ly$\;\alpha$\ $\lambda1216 $ & 4587.9 & 0.6 & 2700 & 200  & 1.0 & 0.2  \\
                      & \multicolumn{2}{c}{ } & \ion{C}{iv} $\lambda1549   $ & 5841.4 & 1.3 & 2850 & 300 & 0.44  & 0.1  \\
                      & \multicolumn{2}{c}{ } & \ion{He}{ii} $\lambda1640  $ & 6184.1 & 0.9 & 2580 & 200  & 0.4 & 0.1 \\
                      & \multicolumn{2}{c}{ } & \ion{C}{iii}] $\lambda1909 $ & 7201.6 & 0.7 & 1260 & 140  & 0.29  & 0.07 \\
NVSS J215226$-$341606 & 1.277 & 0.002         & \ion{Mg}{ii} $\lambda2798  $ & 6383.2 & 0.4 &  280 &  100     & 2.7   & 1 \\
                      & \multicolumn{2}{c}{ } & [\ion{Ne}{v}] $\lambda3426 $ & 7789.1 & 0.7 &  800 &  200     & 4.3   & 2 \\
                      & \multicolumn{2}{c}{ } & [\ion{O}{ii}] $\lambda3727 $ & 8492.0 & 0.9 &  970 &  300     & 6.0   & 3 \\
NVSS J215455$-$363006 & 1.235 & 0.001         & [\ion{Ne}{v}] $\lambda3426 $ & 7659.1 & 0.1 & \multicolumn{2}{c}{$<410$}  & 0.85  & 0.4 \\
                      & \multicolumn{2}{c}{ } & [\ion{O}{ii}] $\lambda3727 $ & 8323.3 & 0.4 &  480 &  100     & 0.89  & 0.4 \\
NVSS J223101$-$353227 & 1.834 & 0.006         & \ion{C}{iv} $\lambda1549   $ & 4400.7 & 0.6 & 1060 & 200       & 0.34  & 0.08 \\
                      & \multicolumn{2}{c}{ } & \ion{He}{ii} $\lambda1640  $ & 4644.3 & 1.5 & 1300 & 500      & 0.14  & 0.03 \\
                      & \multicolumn{2}{c}{ } & \ion{C}{iii}] $\lambda1909 $ & 5402.2 & 0.8 & 2060 & 200       & 0.6   & 0.1  \\
                      & \multicolumn{2}{c}{ } & \ion{C}{ii}] $\lambda2326  $ & 6580.6 & 1.4 & 3690 & 300      & 1.0   & 0.2  \\
                      & \multicolumn{2}{c}{ } & [\ion{O}{ii}] $\lambda2470 $ & 6995.8 & 0.6 & 1070 & 120       & 0.30  & 0.07 \\
                      & \multicolumn{2}{c}{ } & \ion{Mg}{ii}] $\lambda2798 $ & 7945.5 & 0.9 & 2850 & 200       & 1.3   & 0.3  \\
NVSS J223111$-$371459 & 1.248 & 0.002         & [\ion{O}{ii}] $\lambda3727 $ & 8380.7 & 0.4 &  790 &  100     & 11.2  & 3 \\
NVSS J223305$-$365658 & 0.939 & 0.002         & \ion{C}{ii}] $\lambda2326  $ & 4519.2 & 1.3 & 1650 & 800      & 0.59  & 0.3 \\
                      & \multicolumn{2}{c}{ } & [\ion{Ne}{iv}] $\lambda2423$ & 4691.9 & 3.3 & 2700 & 2000     & 0.67  & 0.3 \\
                      & \multicolumn{2}{c}{ } & [\ion{Ne}{v}] $\lambda3426$  & 6643.5 & 0.6 &  670 &  200     & 0.28  & 0.1 \\
                      & \multicolumn{2}{c}{ } & [\ion{O}{ii}] $\lambda3727$  & 7222.2 & 0.1 &  680 &  50      & 1.39  & 0.7 \\
                      & \multicolumn{2}{c}{ } & [\ion{Ne}{iii}] $\lambda3869$& 7500.9 & 0.4 &  520 &  100     & 0.40  & 0.2 \\
NVSS J230004$-$304711 & 0.5445 & 0.0006         & [\ion{O}{ii}] $\lambda3727$  & 5755.3 & 0.8 & 1340 &  400   & 7.5   & 4 \\
NVSS J232125$-$375829 & 2.204 & 0.006         & \ion{C}{iv} $\lambda1549  $  & 4972.8 & 0.5 & 2970 &  300     & 42  & 10 \\
                      & \multicolumn{2}{c}{ } & \ion{C}{iii}] $\lambda1909$  & 6103.4 & 4.0 & 5000 & 2000     & 11.5  & 3 \\
                      & \multicolumn{2}{c}{ } & \ion{Mg}{ii} $\lambda2798 $  & 8975.8 & 0.8 &  830 &  200     &  5.1  & 2 \\
NVSS J233034$-$330009 & 2.675 & 0.005         & Ly$\;\alpha$\ $\lambda1216$  & 4461.7 & 0.3 &  960 &  200     & 4.4   & 2 \\
                      & \multicolumn{2}{c}{ } & \ion{C}{iii}] $\lambda1909$ ? & 5687.8 & 0.2 & \multicolumn{2}{c}{$<550$}  & 0.52  & 0.3 \\
                      & \multicolumn{2}{c}{ } & \ion{C}{iv} $\lambda1549$ ?  & 7017.2 & 3.8 & 4200 & 1500     & 3.9   & 2 \\
NVSS J233226$-$363423 & 0.988 & 0.003         & \ion{Mg}{ii} $\lambda2798$   & 5576.1 &10 & 10700& 500        & 19.2  & 5 \\
                      & \multicolumn{2}{c}{ } & [\ion{O}{ii}] $\lambda3727$  & 7400.1 & 1.3 &  2000&  500     & 11.2  & 3 \\

\hline
\end{tabular}
\end{table*}

Thirty-three per cent of the redshifts obtained are in the range $2<z<3.5$, 
corresponding to
the epoch in which the star formation rate density of the Universe
was at a maximum (see Section~\ref{Intro}). These objects will form part of our
sample for follow-up observations to look at the clustering properties and environments 
of massive galaxies at that epoch.

Three of the possible identifications turned out to be M-stars. These are assumed to be
misidentifications with foreground
stars rather than radio stars, as the surface density of radio stars from \citet{Hel99} 
is much lower than our observed surface density of M-star radio identifications. 
The fraction of contaminating stars is similar to the 
SUMSS--NVSS sample in which three out of 35 spectra were misidentified stars. The
stars are not included in our analysis.
Twelve sources (23 per cent) showed a continuous spectrum with no emission lines and one was not detected,
leaving 25 per cent of the galaxy spectra with undetermined redshifts.
This is a smaller fraction than in the SUMSS--NVSS sample and the
USS sample of \citet{deB01} which found 28 per cent and 35 per cent respectively.
Of our twelve continuum sources, one was a spectrum taken with the ANU 2.3-m in which the S/N was insufficient to
detect any lines and will be reobserved on a larger telescope. Six 
were NTT spectra which may need to be repeated on the VLT to improve
the signal-to-noise ratio, and the remaining five were VLT spectra. 
The `redshift
desert' is the region $1<z<2.3$ where the strong \oii\  line has been
shifted out of the spectral window, but Ly$\alpha$ has not yet moved
into the window.
The continuum sources may be in this redshift range
or, alternatively, the Ly$\alpha$ line may be self-absorbed and the other
emission lines too weak to be detected.

\begin{figure*}
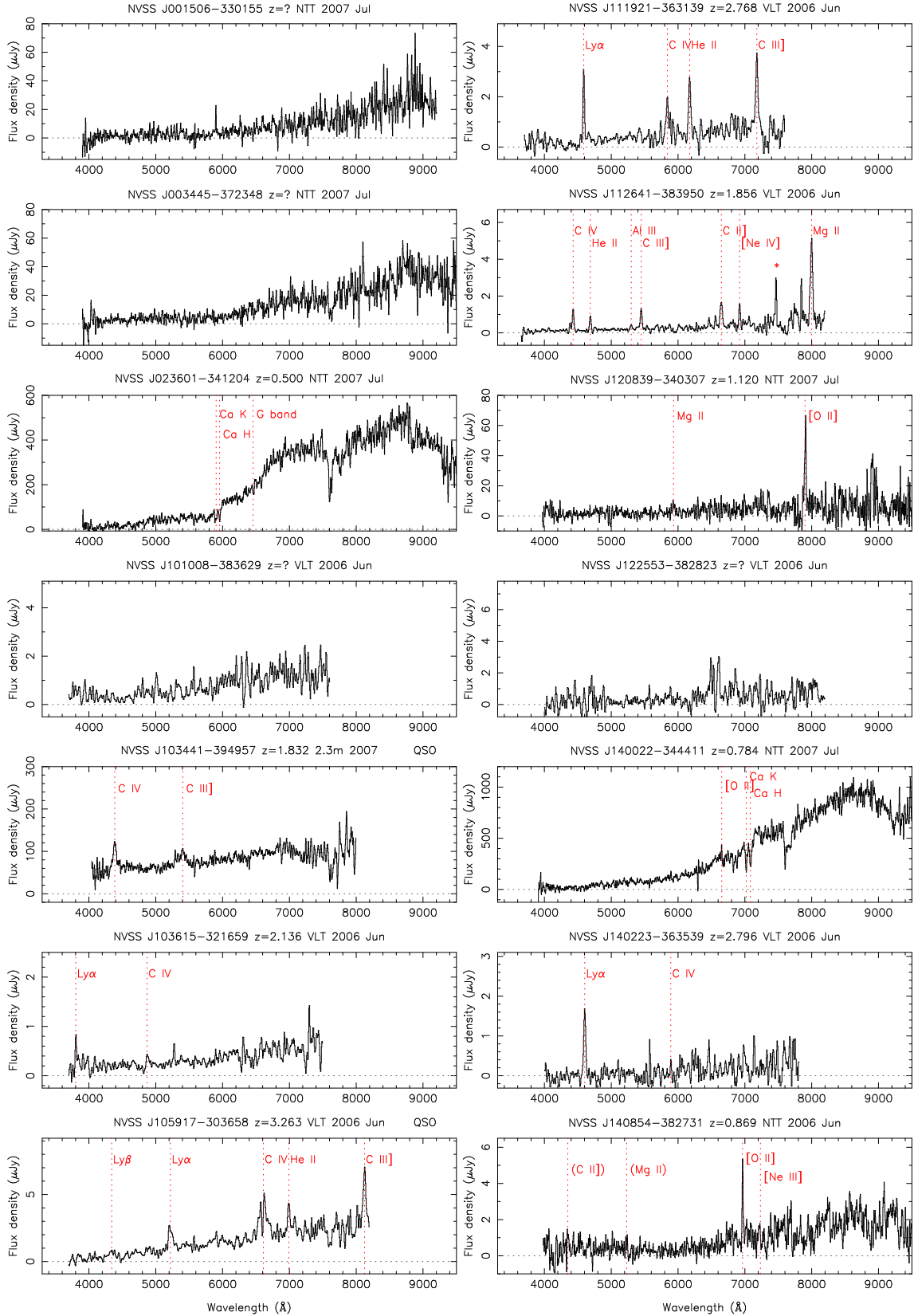

     \centerline{
       \psfig{figure=allspec1fnu.ps,width=7.5cm,clip=t}
       \psfig{figure=allspec3fnu.ps,width=7.5cm,clip=t}
       }
     \vspace{2mm}
     \centerline{
       \psfig{figure=allspec2fnu.ps,width=7.5cm,clip=t}
       \psfig{figure=allspec4fnu.ps,width=7.5cm,clip=t}
       }
 \caption{Spectra of sources with prominent features indicated. The
   source name, redshift and observing run are shown on top of each
   spectrum. 
The zero level is indicated by the
   horizontal dotted line. The positions of the vertical dotted lines
   indicate the predicted wavelengths of the lines at the
   redshift shown and not the wavelength of the fitted peak. Parentheses mark 
marginal detections of lines at the measured redshift and CR represents a cosmic ray. An asterisk shows the 
$\lambda$1305 \ion{O}{i}+\ion{Si}{ii} line appearing in second order.}\label{fig:spectra}
\end{figure*}

\begin{figure*}
     \centerline{
       \psfig{figure=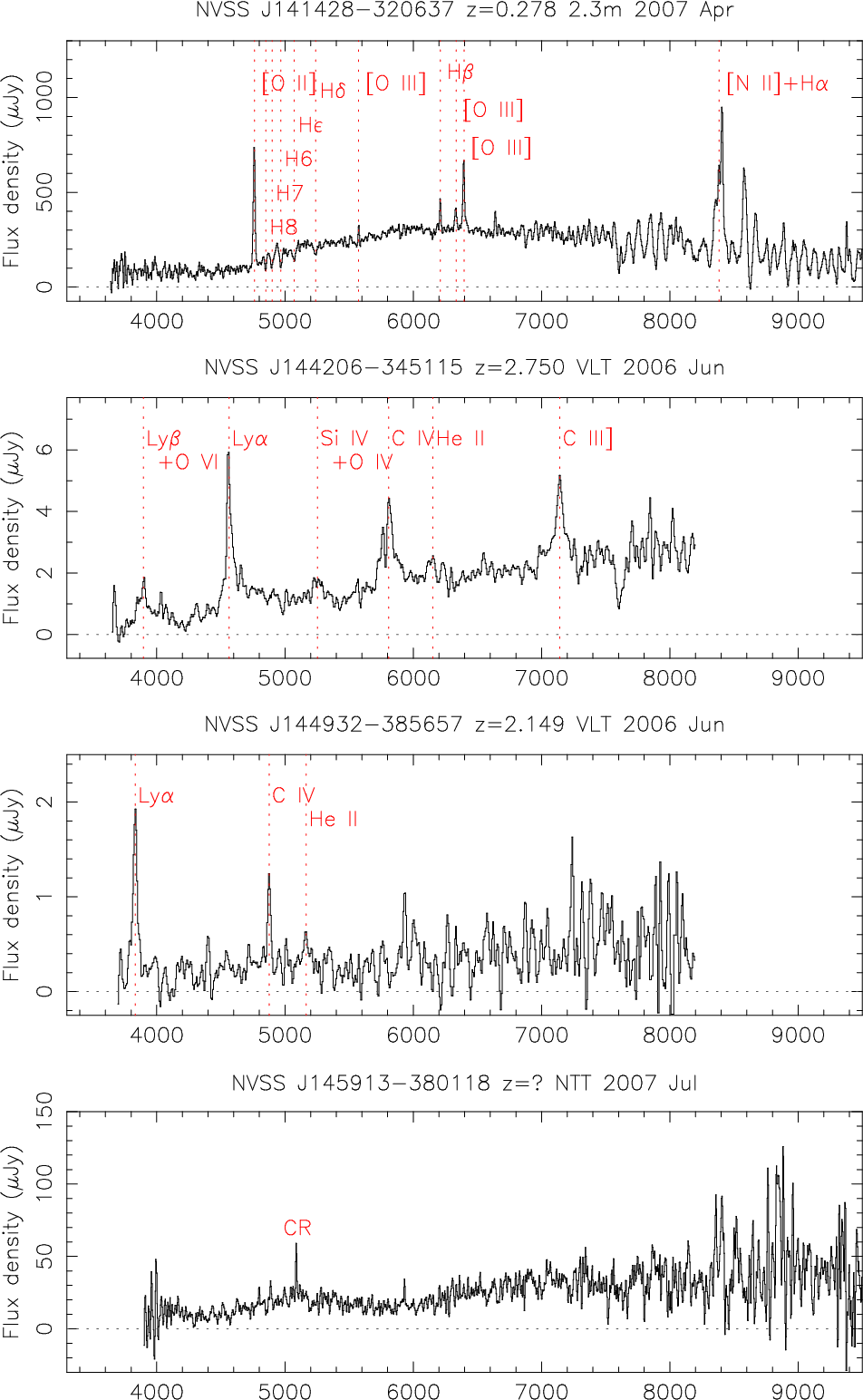,width=7.7cm,clip=t}
       \psfig{figure=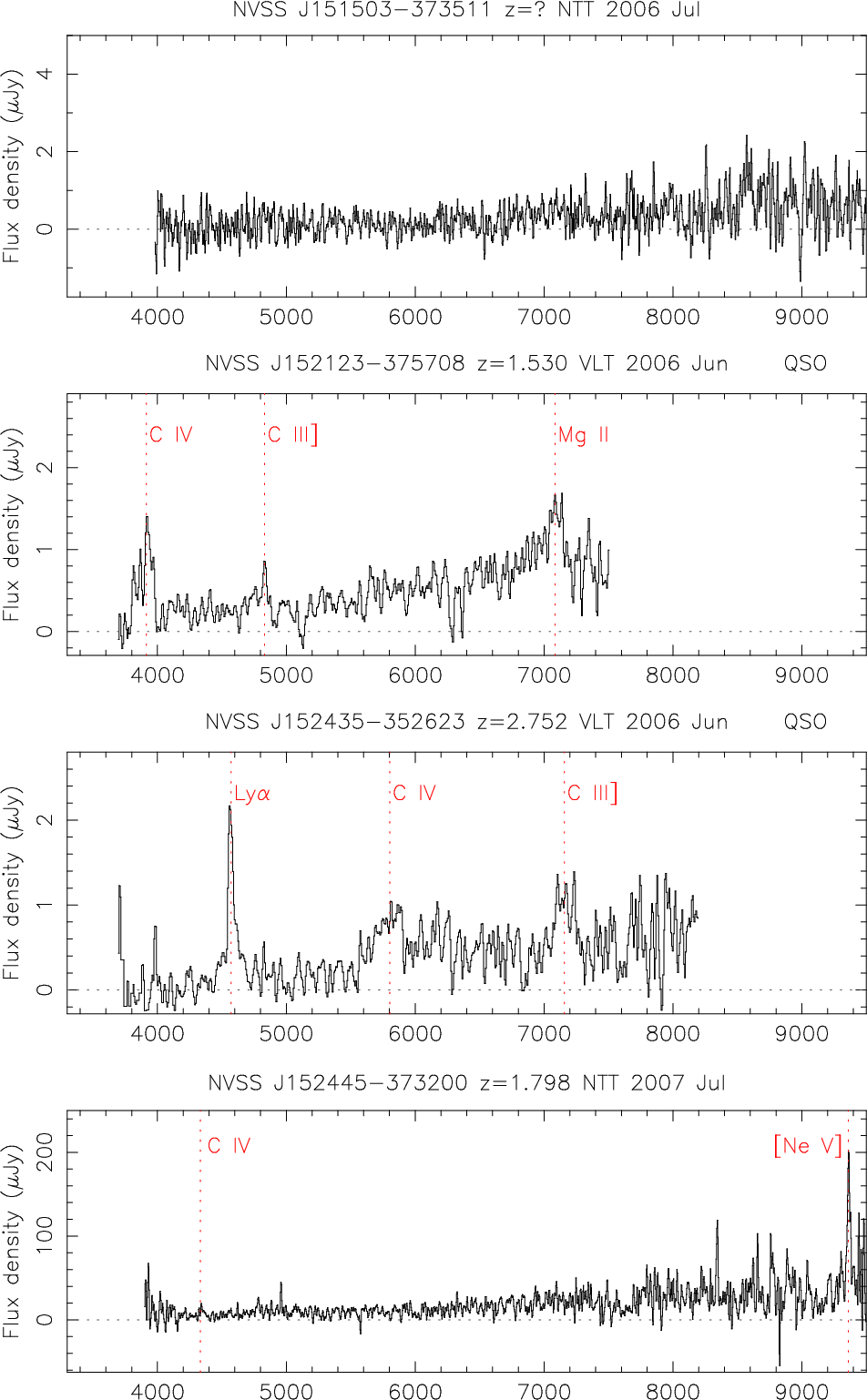,width=7.7cm,clip=t}
       }
     \vspace{2mm}
     \centerline{
       \psfig{figure=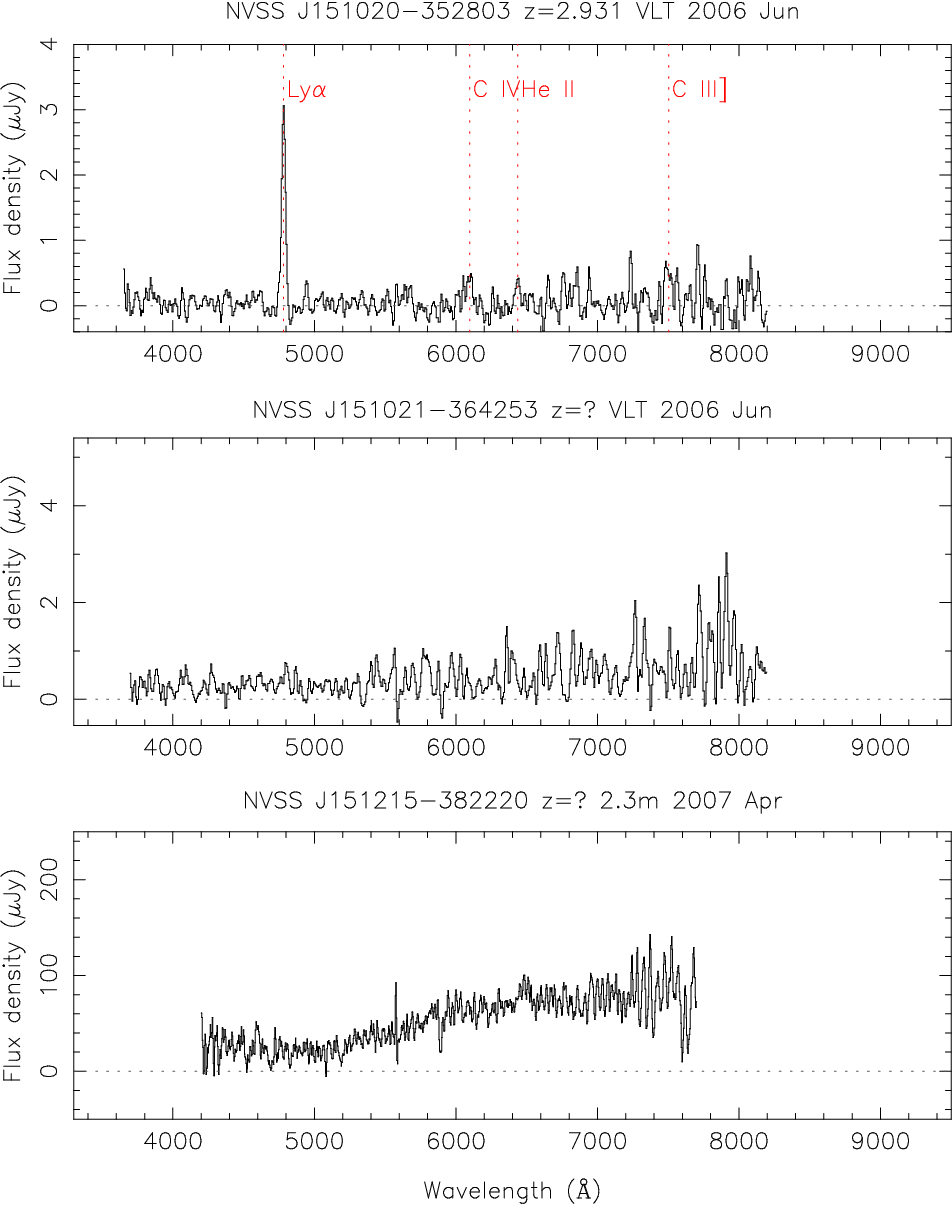,width=7.7cm,clip=t}
       \psfig{figure=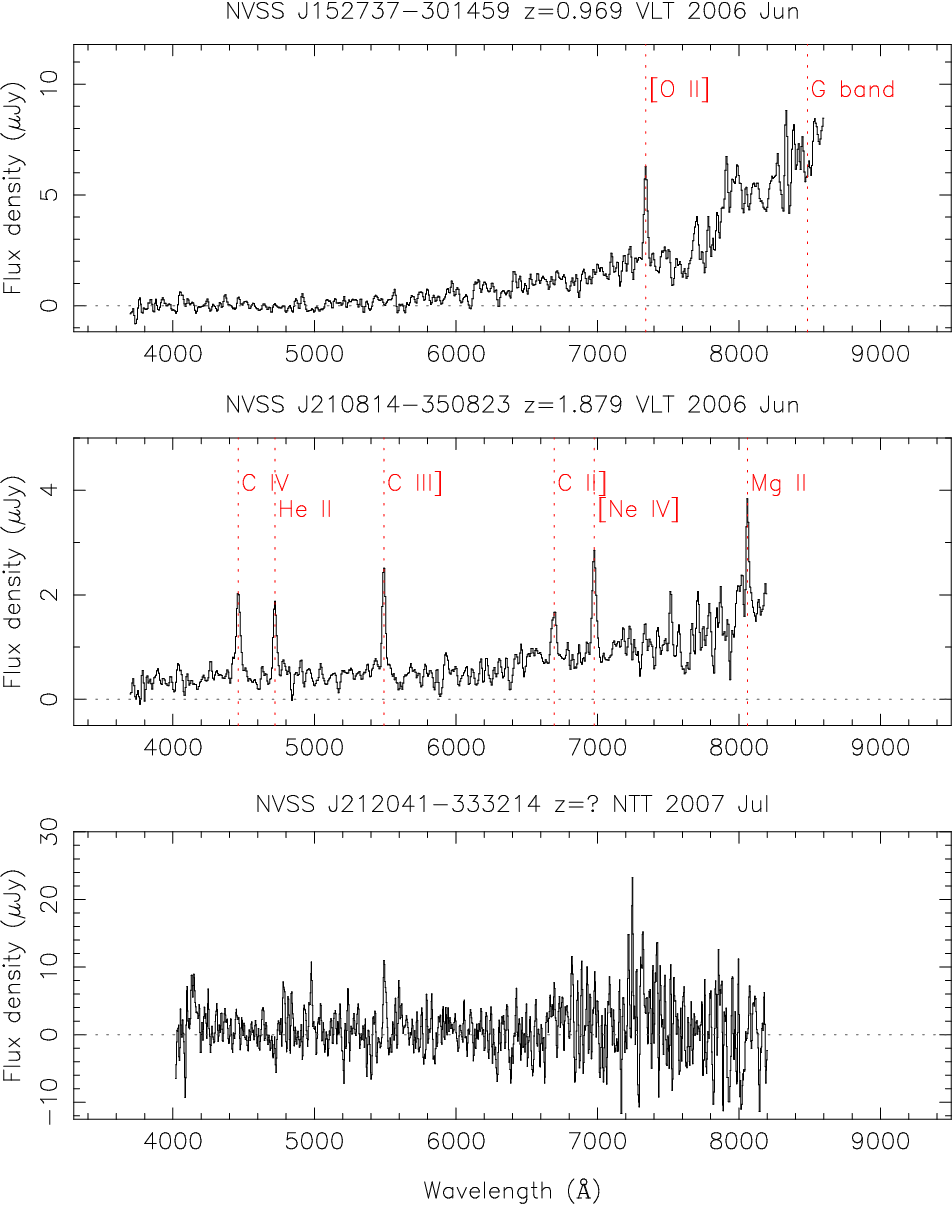,width=7.7cm,clip=t}
       }
     \contcaption{}
\end{figure*}

\begin{figure*}
     \centerline{
       \psfig{figure=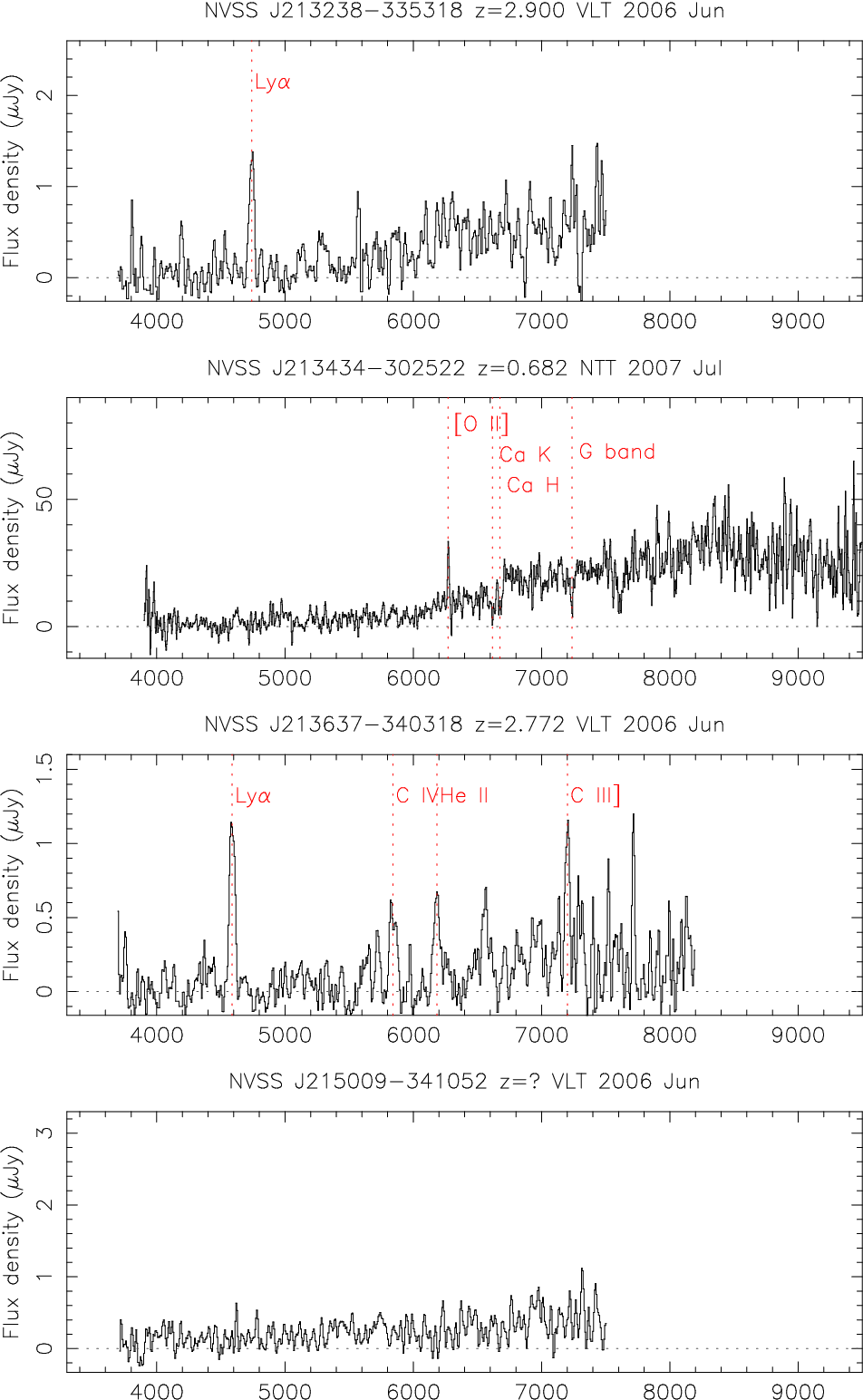,width=7.7cm,clip=t}
       \psfig{figure=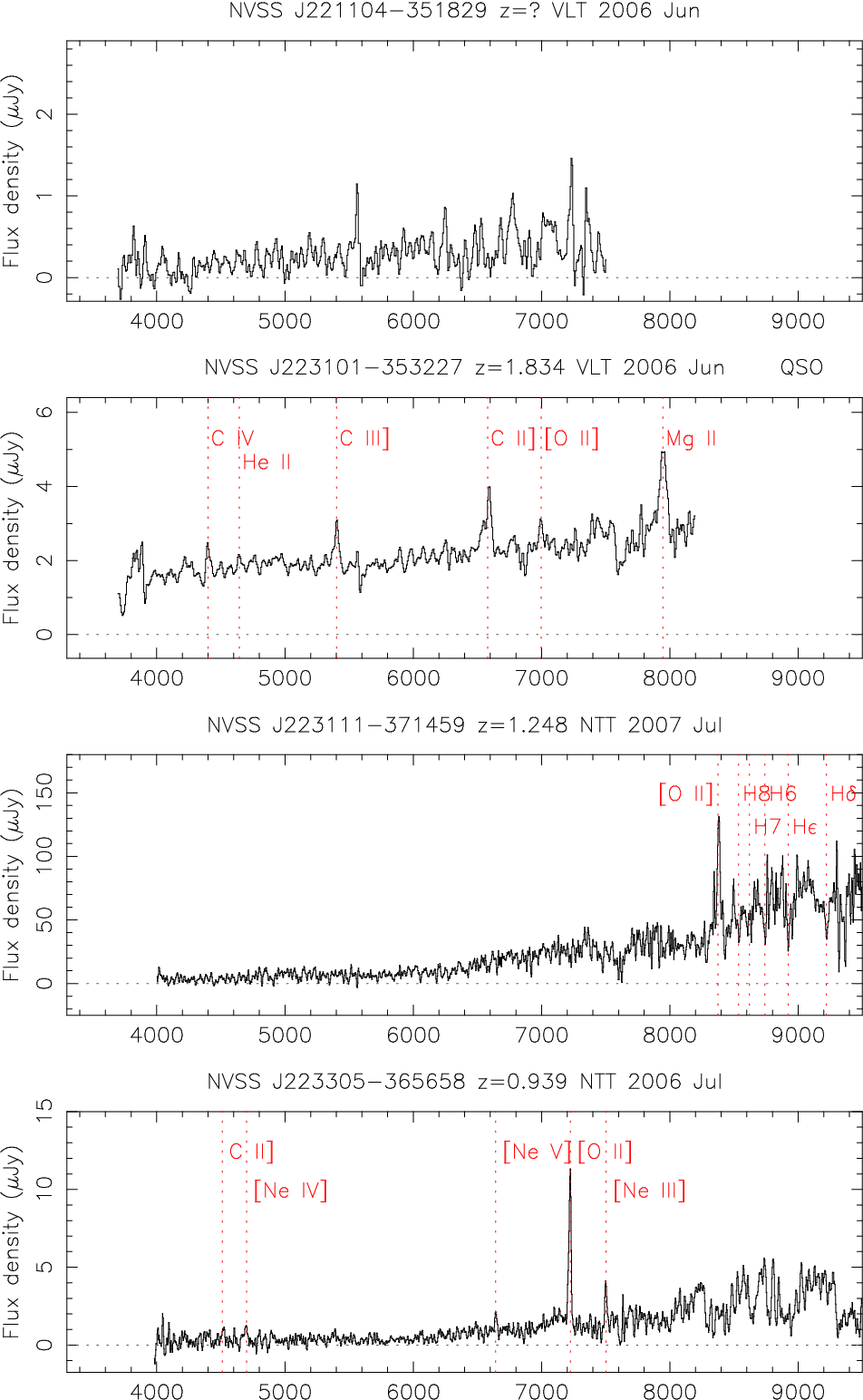,width=7.7cm,clip=t}
       }
     \vspace{2mm}
     \centerline{
       \psfig{figure=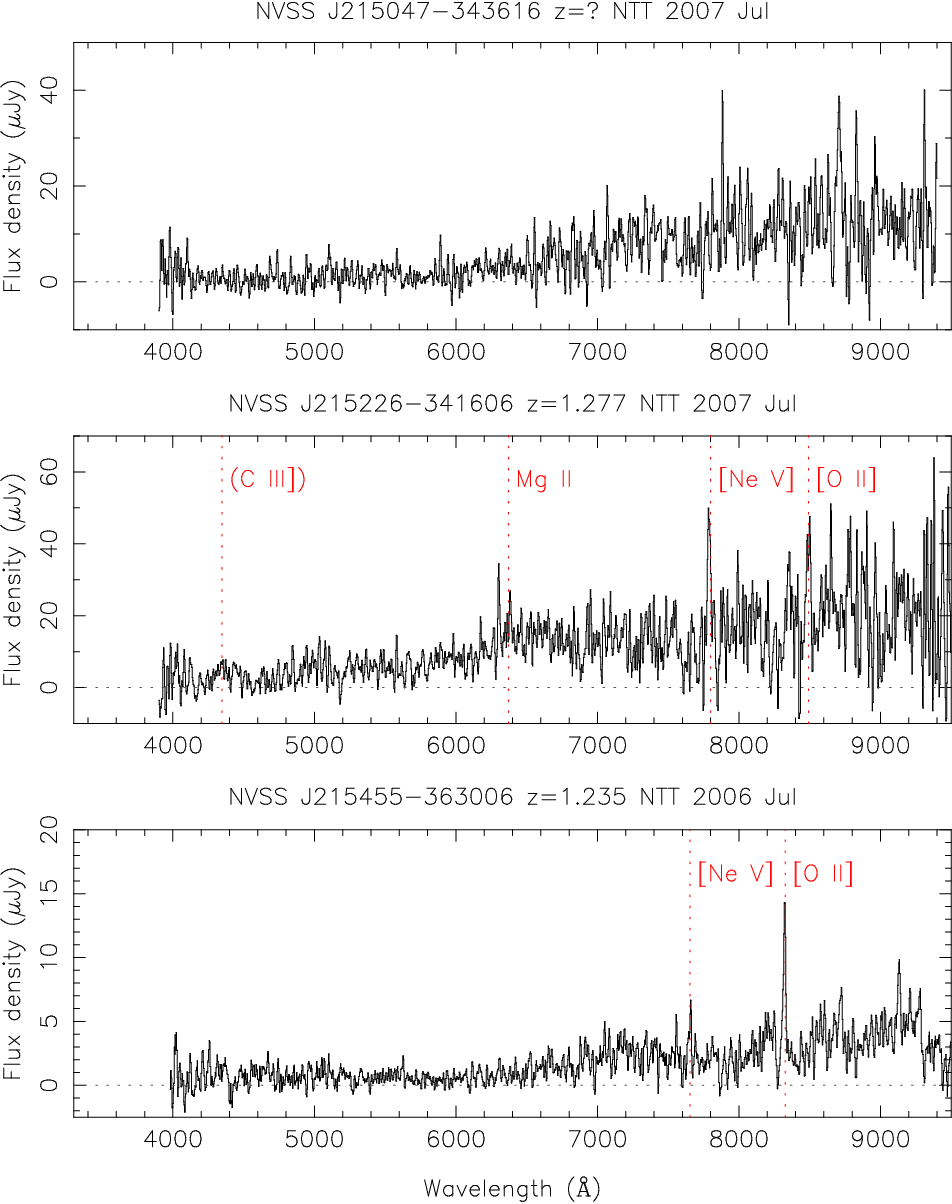,width=7.7cm,clip=t}
       \psfig{figure=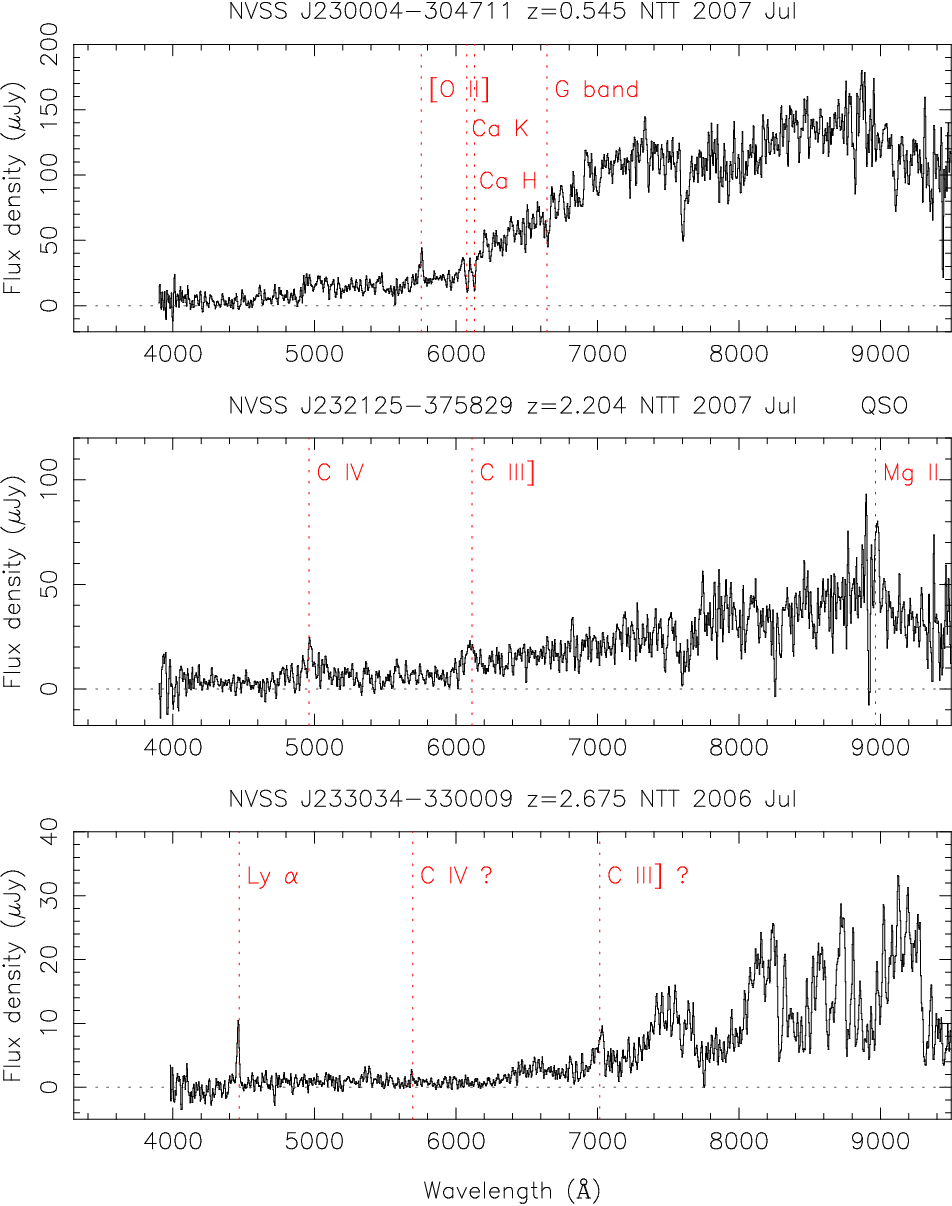,width=7.7cm,clip=t}
       }
     \contcaption{}
\end{figure*}

\begin{figure*}
     \centerline{
       \psfig{figure=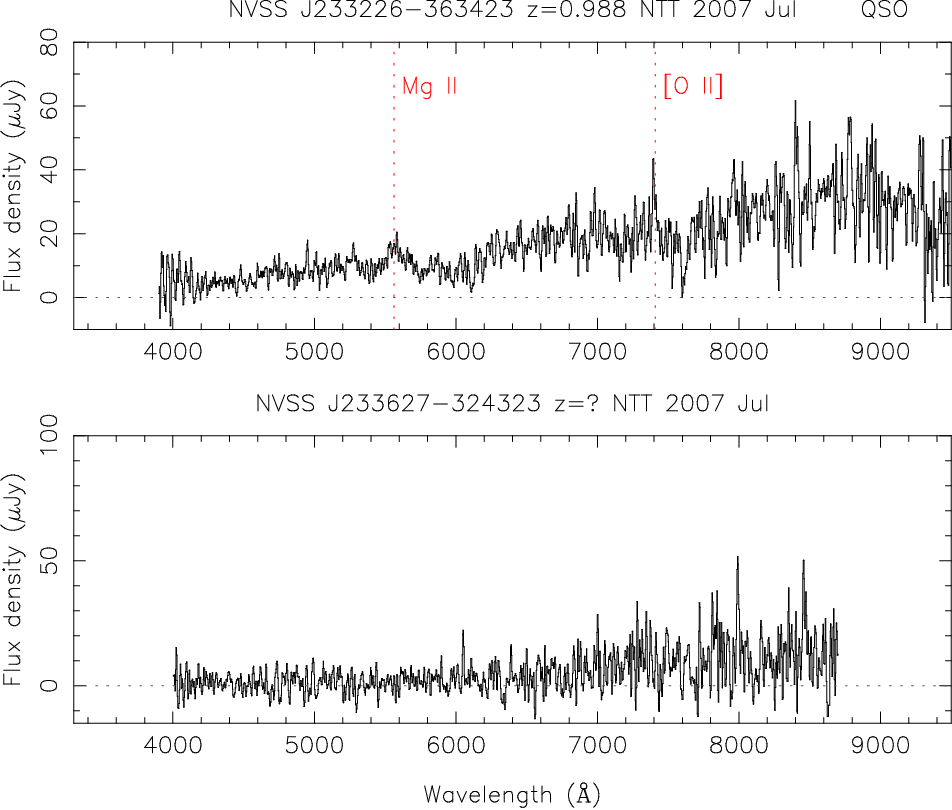,width=7.7cm,clip=t}
       }
     \contcaption{}
\end{figure*}

\subsection{Notes on individual sources}
\label{notes}

\noindent {\bf NVSS~J023601$-$314204/J023605$-$314235 \citepalias[combined source,
see][]{Bry08}:}
The spectrum shows no emission lines, but the redshift is secure based on
stellar features.

\noindent {\bf NVSS~J103441$-$394957:}
The spectrum shows broad QSO emission lines and the radio source has a large linear size.

\noindent {\bf NVSS~J103615$-$321659:} 
We proposed in \citetalias{Bry08} that this may be a double source with very
unequal lobes. While our higher resolution images did not show a double
structure as the fainter lobe was resolved out, our spectrum is that of a
$z=2.136$ galaxy. This high redshift supports our identification as it
would be very unlikely for a $z>2$ galaxy to coincidentally be as close as 
this to a USS radio source. 

\noindent {\bf NVSS~J105917$-$303658:}
The emission lines are broad, typical of a quasar and it has
compact radio structure with a largest angular size of 1.0\,arcsec. The \civ\ line shows strong self-absorption and the blue wing of the
Ly$\alpha$\ line appears to have been absorbed completely (see Fig.~\ref{absorption}). Absorption at the
redshift of the host galaxy was found by
\citet{bha+02} in almost all compact steep-spectrum radio-loud quasars.

\noindent {\bf NVSS~J112641$-$383950:} The line at 7468\AA; 
(marked with an asterisk in Fig.~\ref{fig:spectra})
is the $\lambda$1305 \ion{O}{i}+\ion{Si}{ii} line appearing in second order.

\noindent {\bf NVSS~J140223$-$363539:} 
The 8640-MHz image in \citetalias{Bry08} showed contours extending towards an object
2\,arcsec south-west of the 2368-MHz peak.
A spectrum of this object has confirmed the redshift
to be 2.796. While it does not align with the radio structure, it would be highly
unlikely that this could be a separate high-redshift object that coincidentally
is near the radio source. The NVSS centroid 
is 2.5-arcsec north of the claimed identification. That may indicate the presence of
another faint lobe to the north which is resolved out in the ATCA observations,
suggesting that
it could have a wide-angled tail structure. 
For the time being we assume the
high-redshift source to be the identification due to its redshift, but we
cannot explain the offset from the expected position. 

\noindent {\bf NVSS~J140854$-$382731:} 
The redshift is based on a single line, which we identify as \oii, and
the marginal presence of other lines in the spectrum.

\noindent {\bf NVSS~J141428$-$320637:}
Shows evidence of 
recent star formation, with H$\delta$ absorption characteristic of 
an A-star population.

\noindent {\bf NVSS~J144206$-$345115:}
A quasar, with broad lines and associated absorption (see Fig.~\ref{absorption}).

\noindent {\bf NVSS~J151020$-$352803:}
There is only one bright line in the spectrum, which is spatially extended
over 3.6\,arcsec, which we identify as
Ly$\alpha$, based on the marginal presence of confirming lines.

\noindent {\bf NVSS~J151215$-$382220:} 
No lines were identified, but the brightness of 
the object and the presence of a significant continuum make it 
unlikely to be at high redshift.

\noindent {\bf NVSS~J152123$-$375708:}
A QSO with broad \civ.

\noindent {\bf NVSS~J152435$-$352623:}
Three broad lines including a very broad \civ\ line of 15800\,km\,s$^{-1}$
FWHM confirm that this is a QSO. 

\noindent {\bf NVSS~J213238$-$335318:}
The redshift is based on a single emission line, which we identify
as Ly$\alpha$, based on the absence of 
blueward continuum emission. The bright line is spatially extended over
1.8\,arcsec, supporting the Ly$\alpha$ identification.


\noindent {\bf NVSS~J223101$-$353227:}
The \cii\ and \mgii\ lines are broad, so we classify this object as a QSO.
The line ratios support this interpretation (see section~\ref{Speclinediag}), though 
the line equivalent widths are unusually small 
for a QSO.

\noindent {\bf NVSS~J223111$-$371459:}
The spectrum shows both \oii\ emission and strong Balmer absorption
indicative of an A-star population.

\noindent {\bf NVSS~J232125$-$375829:}
A quasar. The blue end of the spectrum is noisy and has been truncated but Ly$\alpha$ is not detected.

\begin{figure}
\psfig{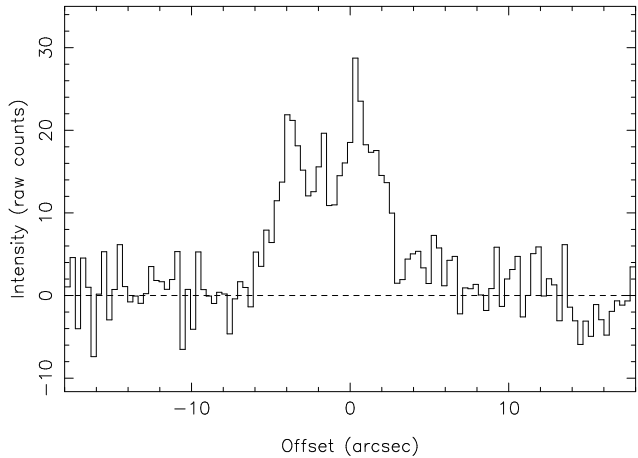}
\caption{Spatial profile (north-south, positive offsets to the north) of the  
Ly$\alpha$ line in the NTT spectrum of NVSS~J233034$-$330009. Ly$\alpha$
covers a FWHM of 7.4\,arcsec or 60\,kpc.} 
\label{Lya}
\end{figure}

\begin{figure}
\psfig{file=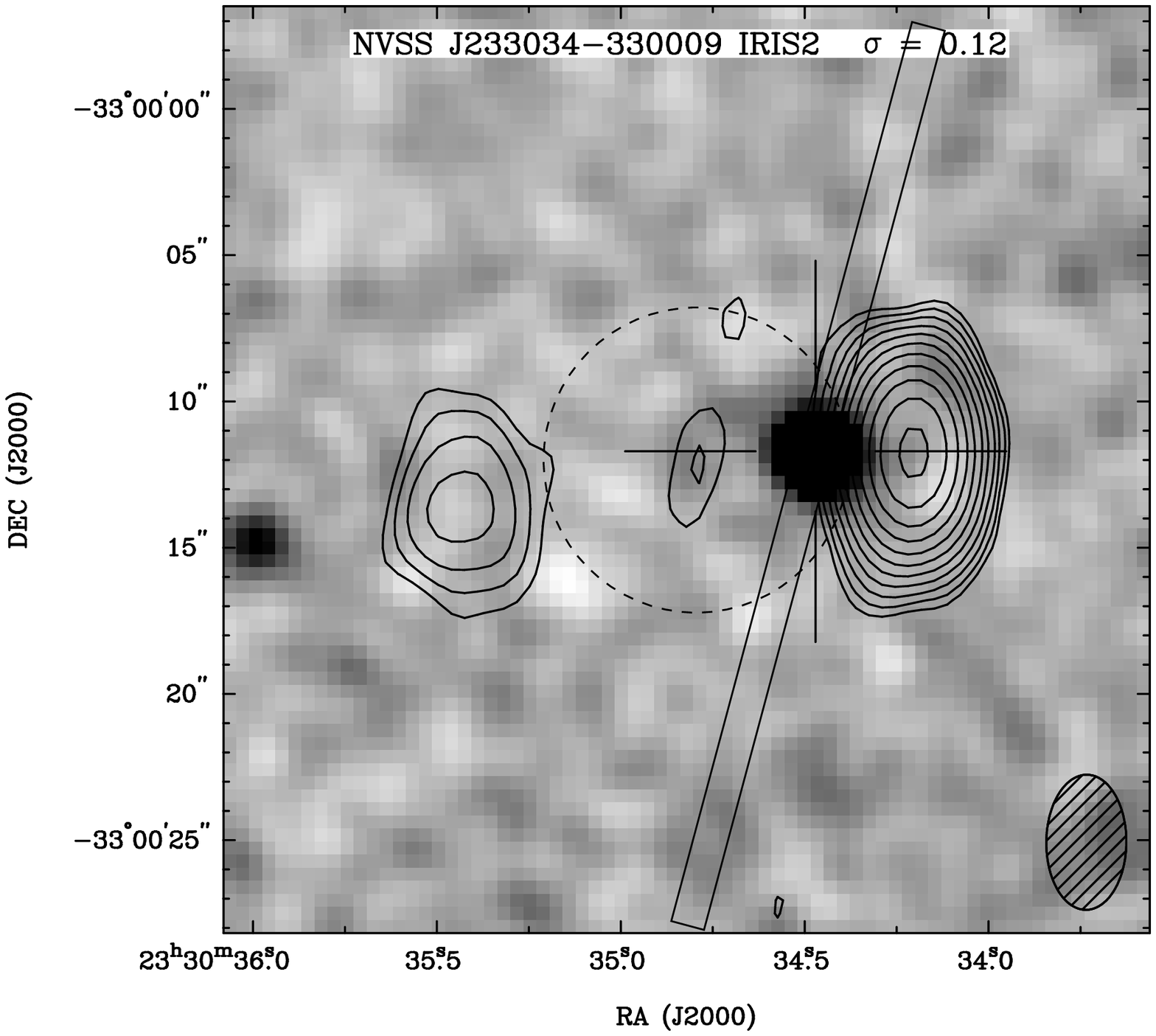, width=8.0cm}
\caption{Our proposed model for NVSS~J233034$-$330009 (see
Section~\ref{notes}) drawn on the overlay of the 4800-MHz ATCA contours on the 
smoothed $K$-band image. The lowest contour is 3 sigma, and the contours
are a geometric progression in $\sqrt 2$. The rms noise ($\sigma$) is shown in the header
in mJy\,beam$^{-1}$.
The parallel lines represent the slit orientation (position angle of 165$^{\circ}$) through
the bright source (with cross-hairs), assumed to be a star. The dashed circle is the
proposed Ly$\alpha$ halo, centred on the radio core, and with a
diameter chosen to reproduce the
observed width of the   
Ly$\alpha$ emission at the slit (7.4\,arcsec), giving a halo diameter
of 10.5\,arcsec or 85\,kpc.} 
\label{Lya_halo}
\end{figure}

\noindent {\bf NVSS~J233034$-$330009:}
From our original 1384- and 2368-MHz radio images, we initially chose the 
identification to be the bright $K$-band object between the lobes and closer to 
the stronger lobe. We had taken a spectrum of this object
before the 4800-MHz-high-resolution image \citepalias[shown in][]{Bry08} 
identified a possible core which
is 4-arcsec east of the bright $K$-band object. The spectrum is typical of an M-star,
but there is a clear emission line in the blue which we identify as Ly$\alpha$ at a redshift of 2.675. The slit position
angle was $165^{\circ}$ and centred on the bright object and therefore does not 
go through the core position.
The spatial profile of the Ly$\alpha$ line, shown in Fig.~\ref{Lya}, is 
extended with a FWHM of 7.4\,arcsec (60\,kpc)
along the slit. We therefore believe that the correct 
$K$-band identification is fainter than the limit on the IRIS2 image (i.e. $K>19$)
and that it is surrounded by a Ly$\alpha$ halo which extends from the core to beyond 
the bright source,
which is an M star. In Fig.~\ref{Lya_halo}, we show the possible
position of the Ly$\alpha$ halo. In a simplistic picture, where the Ly$\alpha$ 
halo is spherical, the 7.4-arcsec FWHM of the emission line at the slit can be
reproduced by a halo of diameter $\sim10.5$\,arcsec, or 85\,kpc.
With an alternate geometry, the halo size would need to be a minimum of
32\,kpc, the
projected distance from the core to the bright object. This is a modest
size for a Ly$\alpha$ halo, as some have been found to be $>100$\,kpc 
\citep{Kur00a,Ven:02,reu:07}.
Since the possible core was not detected in our
8640-MHz image this interpretation is open to question. 

An alternative picture is that the bright $K$-band object is a galaxy and
star coincident along the line of sight, and the middle component 
is a hotspot along the eastern jet. 
If this picture is correct then 
there is a marginal line at the wavelength expected for \civ. 
A further line sits at the expected position of \ciii, but it cannot be 
differentiated from the M-star features at that wavelength.


\noindent {\bf NVSS~J233226$-$363423:}
A very broad \mgii\ line, with a FWHM of $10^{4}$\,km\,s$^{-1}$ indicative of a QSO.

\section{Discussion}

\subsection{$K$--$z$ distribution}

\subsubsection{Emission-line contribution to $K$-band magnitudes}
Strong emission lines such as \oii\ , \oiii\ and
H$\alpha$\,$\lambda$6562 can be shifted into the $K$-band, particularly
at redshifts beyond $\sim2$. The contribution of these
lines to the $K$-band magnitude is primarily dependent on redshift,
but also on radio luminosity, due to the correlation between
emission-line and radio luminosities \citep{Wil99}.
\citet{jar01} calculated the emission-line contamination at each
redshift for different 151-MHz flux densities.
Adopting their approach, 
we have estimated the emission-line contamination to our $K$-band
magnitudes for the MRCR--SUMSS sources. 
Firstly, we extrapolated our
radio spectrum to 151\,MHz, using the spectral index from our 5-point fit or,
when unavailable, using the 843--1400-MHz spectral index \citepalias[see][]{Bry08}.
Some of the sources have a measured 151-MHz flux density (see Table~\ref{151})
which was used. We then estimated the
$K$-band contribution from emission lines using the information from \citet[][fig.2]{jar01}, 
and then subtracted the emission-line 
contribution from the measured $K$ flux. We find that in all but one case 
(NVSS~J144932$-$385657) the
typical contribution of the emission lines
is less than the errors on our $K$ magnitudes. 
The inherent assumptions in this process mean that a typical 
emission-line contamination was calculated, rather than an actual emission-line
contribution. These assumptions include using average line ratios, the 
generalisation of the correlation between emission-line
and radio luminosity, and the assumption that our radio
spectra remain straight down to 151\,MHz. 

The only source where line emission may contribute to the $K$ flux is
NVSS~J144932$-$385657 at
$z=2.149$, where the emission lines of \nii\,  and H$\alpha$ 
fall in the $K$ band. This source is relatively faint at $K=19.8\pm0.2$ and 
the possible emission-line contribution may account for up to 
0.4\, magnitudes of $K$-band flux. 
However, the spectrum in Fig.~\ref{fig:spectra} shows that 
the emission lines are weak
and therefore the lines that are shifted into the $K$-band may not
have the maximum flux calculated here.
Furthermore, this object is not an outlier in the $K$--$z$ distribution (see Fig.~\ref{Kz} below) so there is no evidence that
emission lines have boosted its $K$ magnitude.

\subsubsection{Trends on the $K$--$z$ correlation}
\label{Kztrends}
\begin{figure*}
\centerline{\psfig{file=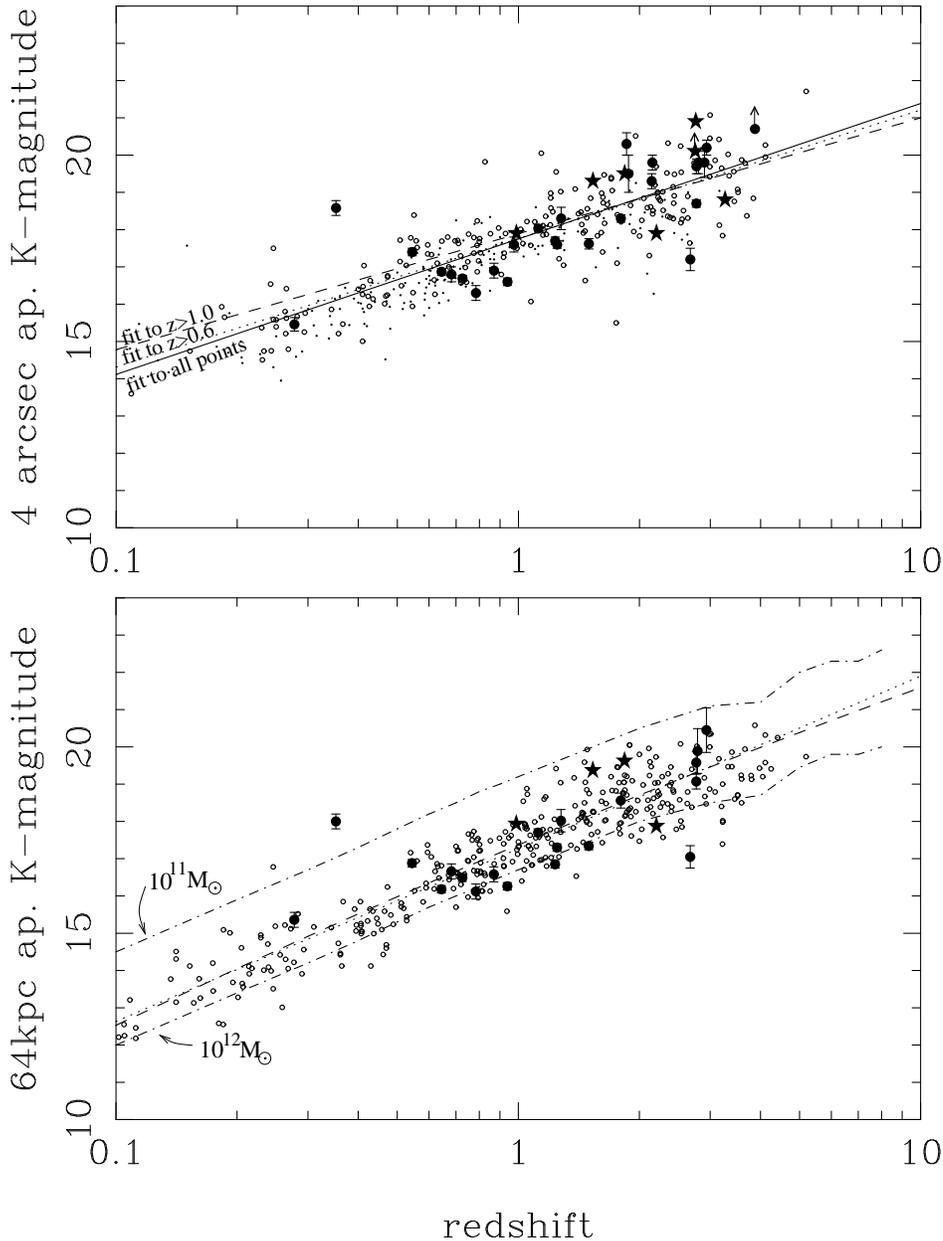,width=12.4cm}}
 \caption{{\it Top}: $K$ magnitudes in 4-arcsec apertures versus redshift
for the MRCR--SUMSS HzRGs (filled circles), the MRCR--SUMSS QSOs (stars)
and samples from the literature (small open circles). The literature
samples were chosen because they had 4-arcsec $K$-band and redshift
information. QSOs, uncertain redshifts, and lower limit $K$ magnitudes
were removed from the literature samples. Two of the samples 
(CENSORS, \citeauthor{bro06} \citeyear{bro06}, \citeyear{bro08}; 
and 6C**, \citeauthor{cru06} \citeyear{cru06}) 
had 3- and
5-arcsec-aperture $K$ magnitudes and a 4-arcsec-aperture magnitude was
calculated by averaging the equivalent 3- and 5-arcsec fluxes and
converting to a 4-arcsec magnitude. The other samples plotted are the 6C*
from \citet{jar01}, \citet{deB01,deB02} sample, 7C from
\citet{Wil03}, SUMSS--NVSS from \citet{deB04,deB06} and
the sample from \citet{bor07}. The small dots are the 
\citetalias{McC96} sample 
in which the magnitudes were
measured in a 3-arcsec aperture. Three straight line fits to both the
literature samples and the MRCR--SUMSS sample are shown; a fit to points
at all redshifts (solid line, Eq.~\ref{eq1}), to points with $z>0.6$
(dotted line, Eq.~\ref{eq2}) and to points with $z>1$ (dashed line,
Eq.~\ref{eq3}). The McCarthy points were not used in the fit calculations
due to the smaller measurement aperture for that sample and neither were
the QSOs and objects with contamination by a foreground star in the
MRCR--SUMSS sample.  The redshift errors for our MRCR--SUMSS data are much
smaller than the plotted filled circle as are some of the $K$-magnitude
errors.  {\it Bottom}: $K$ magnitudes in a 64-kpc metric aperture versus
redshift for the MRCR--SUMSS HzRGs (filled circles), the MRCR--SUMSS QSOs
(stars) and samples from the literature (small open circles). The
literature samples are the 3CRR, 6CE, 6C$^{*}$ and 7CRS radio galaxy samples
compiled by \citet{Wil03}, the composite samples of \citet{vanB98} 
and \citet{deB02}, the SUMSS--NVSS sample and the
McCarthy sample. The 64-kpc aperture magnitudes for the MRCR--SUMSS,
SUMSS--NVSS and McCarthy samples were calculated from 8-arcsec-aperture
magnitudes (see text for limitations). We have also plotted the
corresponding 64-kpc and 8-arcsec-aperture fit equations from \citet{Wil03} 
(dashed line) and \citet{deB04} (dotted line)
respectively. The dot-dash lines are the $10^{11}$\mdot and $10^{12}$\mdot
elliptical galaxy evolution models from \citet{Roc04}. }
 \label{Kz}
\end{figure*}

The $K$--$z$ correlation is an empirical relationship that spans redshifts up to
$z>4$. While the correlation remains linear\footnote{Linear fit
between $\log(z)$ and $K$ magnitude corresponds to a power-law
relation between $z$ and $2.2\mu$m flux.}, different galaxy samples have
been shown to give different linear fits to the $K$--$z$ relation. 
\citet{Wil03} made a fit to 64-kpc-aperture magnitudes
based on the 3CRR, 6CE, 7C and 7CRS galaxy samples. The 3CRR sample
was found to be $\sim0.6$ magnitudes brighter in $K$ at $z>1$ than the 6C
\citep{jar01}, while the 7C and 3CRR are offset in $K$-magnitude at
all redshifts \citep{Wil03}. A fit to the $K$--$z$ relation for
FIRST radio sources \citep{ElB07} was also found to be fainter than the
\citet{Wil03} relation by more than 0.5 magnitudes at high
redshift and even more at low redshift. These differences highlight the
selection effects (including flux density limit) in any sample and show that the fits to the $K$--$z$
plot depend on which samples are included and the aperture correction
used. 
An individual fit, therefore, has no physical meaning but serves only to
highlight whether a galaxy lies above or below the trend for that sample.

In Fig.~\ref{Kz} the 4-arcsec-aperture $K$ magnitudes are plotted
against redshift for all the sources in our sample that have spectroscopic
redshifts, along with many samples from the literature with
4-arcsec-aperture $K$ magnitudes. These are both USS-selected samples and
non-USS-selected samples as listed in the Figure, and include the largest
non-USS sample (\citeauthor{McC96} \citeyear{McC96}; 2007, private communication) 
comprising 277
$K$-band observations of radio galaxies, of which 175 have redshifts.  
The advantages of using the 4-arcsec-aperture magnitudes are discussed in
\citetalias{Bry08} and in terms of the $K$--$z$ dispersion in
Section~\ref{sec:dispersion}. We have therefore chosen to adopt a fit to
the 4-arcsec-aperture magnitudes from the combined literature samples as
the most suitable metric that represents our data and allows more of the
galaxies in our sample to be compared with the linear relation for
HzRGs. For consistency with some papers in the literature we also show the
$K$--$z$ plot based on 64-kpc aperture values.  The MRCR--SUMSS 64-kpc 
magnitudes  are calculated from 8-arcsec-aperture magnitudes
following the procedure described by \citet{eal97} and listed in
Table~\ref{data}. 

Three fits to the 4-arcsec-aperture magnitudes in Figure~\ref{Kz} are shown.
 The fits are linear least-squares fits, because in each case the error
on a quadratic fit was higher. QSOs and objects with contamination from
foreground stars were not included in the fits. 
 The equations of the fit lines are 
\begin{equation}
K(4'') = 3.64 \log(z) + 17.75 
\label{eq1} 
\end{equation}
for objects at all redshifts (solid line),
\begin{equation}
K(4'') = 3.45 \log(z) + 17.76 
\label{eq2}
\end{equation}
for $z>0.6$ objects only (dotted line), and
\begin{equation}
K(4'') = 3.11 \log(z) + 17.89 
\label{eq3}
\end{equation}
for $z>1.0$ objects (dashed line).

As the redshift range increases to include lower redshift objects, the fit
steepens, whereas if the 4-arcsec aperture was missing any $K$-band flux
we would expect the opposite effect, i.e., the fit would be flatter. There
is therefore no compelling evidence that the lower redshift galaxies have
$K$-band emission detected beyond a 4-arcsec aperture.

In Fig.~\ref{Kz} we have also plotted the evolutionary curves for
elliptical galaxies with masses of $10^{11}$\mdot and $10^{12}$\mdot from
the models of \citet{Roc04}. The majority of identifications in
our sample are consistent with elliptical galaxies of mass $10^{11} -
10^{12}$\,\mdot. 
\citet{Sey07} modelled HzRG masses from $H-$band
magnitudes using early-type galaxy models from PEGASE2 \citep{fio97} 
with galaxy formation at $z=10$ and a different
initial mass function to that used by \citet{Roc04}. \citet{Sey07} found that
the majority of their galaxies fitted within the $10^{11}$--$10^{11.5}$
\mdot model over a redshift range of $1<z<4$.

The $K$--$z$ distribution has a scatter about the fitted line due to the
variations in stellar content of the galaxies, the amount of dust, and the rest
wavelength range being sampled at each redshift. At high redshift $UV$
light is observed in the near-IR, and this may introduce a contribution
from the AGN continuum. In Fig.~\ref{Kz} there are two galaxies that are
more than $2.5\sigma$ from the adopted best fit line (Eq.~\ref{eq1}), and
lie well outside the $10^{11}$\mdot and $10^{12}$\mdot model lines.  
NVSS~J011606$-$331241 is $3.3\sigma$ fainter than the fit to the main
distribution of points, with $K=18.6$ and $z=0.352$. This galaxy is in
common with the SUMSS--NVSS sample; in \citet{deB06} it was
concluded that there is a chance the correct identification may be a
fainter nearby galaxy but higher resolution radio imaging is needed to
confirm this. The other main outlier is NVSS~J233034$-$330009 which is
$2.8\sigma$ brighter than predicted, with $z=2.675$ and $K=17.2$; this
source is discussed in detail in Section~\ref{notes}.

\subsubsection{Dispersion in the $K$--$z$ distribution}
\label{sec:dispersion}

An increase in the dispersion of points on the $K$--$z$ plot at high redshift
has been used in the past to constrain the epoch of formation of
radio galaxies. \citet{eal97} found an increase in dispersion above 
$z\sim2 $, and concluded that those galaxies were in the process of
formation while those at $z<2$ are evolving passively. 
On the other hand, \citet{jar01} 
found no evidence for an increase in the dispersion of $K$ magnitudes
at $z>2$, and concluded that radio-luminous systems formed most of their
stars at $z>2.5$ and have evolved passively since then.

In Table~\ref{dispersion}, we have calculated the dispersion in redshift bins for all the galaxies
in Fig.~\ref{Kz} based on each fit shown. This shows that the dispersion in the 64-kpc-aperture
points around the \citet{Wil03} fit is no lower than dispersion of the 4-arcsec-aperture points
about the fit to all redshifts. Therefore, we have adopted the 4-arcsec-aperture metric with the fit
in Eq.~\ref{eq1} in order to compare the MRCR--SUMSS galaxies with the trend in radio galaxies for the
rest of this paper. 
There is no evidence for an increase in dispersion above $z=2$ for the 4-arcsec-aperture fits.
This supports the results found by \citet{jar01} and \citet{deB06} that 
radio galaxy hosts have been evolving passively since epochs corresponding to $z=3$ or earlier.

\begin{table}
\caption{The dispersion in the $K$-magnitude as a function of redshift for all the galaxies in Fig.~\ref{Kz} but excluding QSOs and galaxies that are contaminated by foreground stars. 
The number of galaxies in each redshift bin is
given in brackets beside the dispersion.
}
\begin{tabular}{lccc}
\hline
\hline
Fit equation & \multicolumn{3}{c}{Dispersion (mag)} \\
        & all z's & $z<2$ & $z>2$  \\
\hline
64-kpc fit            & 0.7 (534)  & 0.7 (441) & 0.9 (93) \\
all 4$''$ ap.            & 0.7 (211) & 0.7 (146) & 0.7 (65)\\
$z>0.6$ for 4$''$ ap. & 0.7 (211) & 0.8 (146) & 0.7 (65) \\
$z>1.0$ for 4$''$ ap. & 0.8 (211) & 0.8 (146) & 0.7 (65) \\
\hline
\label{dispersion}
\end{tabular}
\end{table}

\subsubsection{QSOs in the MRCR-SUMSS sample}

QSOs are usually removed from HzRG samples on the basis that their
$K$-band flux is generated by the central AGN rather than from stellar
light. The implication is that the non-thermal AGN flux would be stronger
than the thermal contribution from the host and hence bias the $K$--$z$
distribution in favour of brighter $K$-band magnitudes.  The unification
model of AGNs \citep{Bar89,ant93,urr95} 
suggests that the difference between radio galaxies and radio-loud QSOs is
an orientation effect, such that the QSOs are viewed so the nucleus can be
seen directly rather than being obscured by a dusty torus as in a radio
galaxy. QSOs would therefore be expected to have smaller apparent linear
sizes. While \citet{Bar89} found that QSOs, on average, were a factor of
two smaller than radio galaxies, many QSOs have large radio linear sizes,
up to several Mpc (\citeauthor{Ril89} \citeyear{Ril89}; \citeauthor{bha98} \citeyear{bha98}; 
\citetalias{Bry08}, section 3.2.1) 
and some do not show very bright central sources. Alternatively, it has
been argued that the difference between QSOs and radio galaxies may be due
instead to their evolutionary stage and/or environment.  For instance,
compact radio sources were found to live in denser environments than
extended sources by \citet{Pre88} 
and this was used as an
argument for them being different types of sources.  It is possible,
therefore, that radio morphology may depend more on the age or environment
of a source than its viewing angle.  As a result we consider it important
to include QSOs in any follow-up study of HzRG environments.

It has been suggested that the alignment effect between radio structures
and UV/optical light seen in HzRGs is also apparent in QSOs \citep{Leh99} 
and can be explained by scattering of light from the active
nucleus rather than jet-induced star formation.  Scattering models are
favoured because the emission is highly polarised. While broad emission
lines have long been the defining features of QSOs, 
the distinction between radio-loud QSOs and radio galaxies is
not clear-cut.
It is worth noting that compact
steep-spectrum QSOs in the Molonglo Quasar Sample \citep{Bak95} 
also have properties similar to HzRGs:  a heavily reddened optical
continuum, strong narrow forbidden lines and narrower than usual permitted
lines. USS QSOs are quite rare, and therefore may have different physical
characteristics to the bulk of the QSO population.

While our fits to the $K$--$z$ plot do not include the QSOs, it can be
seen in Figure~\ref{Kz} that the QSOs tend to fall on the {\it faint\/}
side of the fit to all redshifts, indicating that the $K$-band flux is not
unusually strong in comparison with the radio galaxies. If the QSOs had
been included, the $z>1$ fit does not change Eq.~\ref{eq3} within the
precision shown.
We note, however, that target selection for spectroscopic followup was based on $K$-band 
magnitude but not on $K$-band morphology; hence bright QSOs will not 
be included in our sample. 

While the objects with line widths $>3000$\,km\,s$^{-1}$
have been classified as QSOs, we find no evidence from the $K$-band
magnitudes that we are seeing the nuclear source directly.  Furthermore,
two of the MRCR--SUMSS QSOs --- NVSS~J223101$-$353227 and
NVSS~J233226$-$363423 --- have extended $K$-band emission that is 
aligned (within 7- and 3-degrees respectively) with the radio
structure.

\begin{figure*}
\centerline{\psfig{file=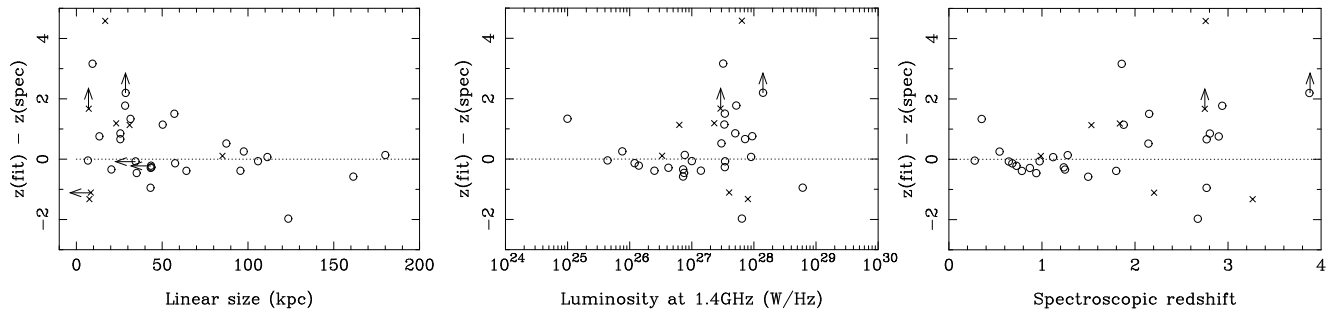, width=17.5cm}}
\caption{The difference between predicted and spectroscopic 
redshifts (where the predicted redshift comes from the $K$--$z$ 
fit in Fig.~\ref{Kz}, Eq~\ref{eq1}) as a function of linear size, 
1.4-GHz luminosity and spectroscopic redshift. Circles are radio galaxies and crosses
mark the QSOs.  }
\label{zz}
\end{figure*}

In Fig.~\ref{zz}, we compare the MRCR--SUMSS radio galaxies and QSOs with
the overall $K$--$z$ fit for radio galaxies defined by 4-arcsec aperture
magnitudes. 
The offsets between the measured redshifts and those predicted
by Equation~\ref{eq1}, $z$(fit)$ - z$(spec), are plotted against radio
luminosity, linear size and redshift.  It is clear that the differences
among the $K$--$z$ fits defined by Eqs.~\ref{eq1}, ~\ref{eq2} and
~\ref{eq3}
are small in comparison to the scatter about each fit.
Moreover, these QSOs do not stand out from the HzRGs in linear size, radio
luminosity or redshift distribution. It is {\it only\/} the spectral line
width that differentiates QSOs from HzRGs in the MRCR--SUMSS sample.

\subsection{Linear size}
\label{linear_size}

\begin{figure}
\centerline{\psfig{file=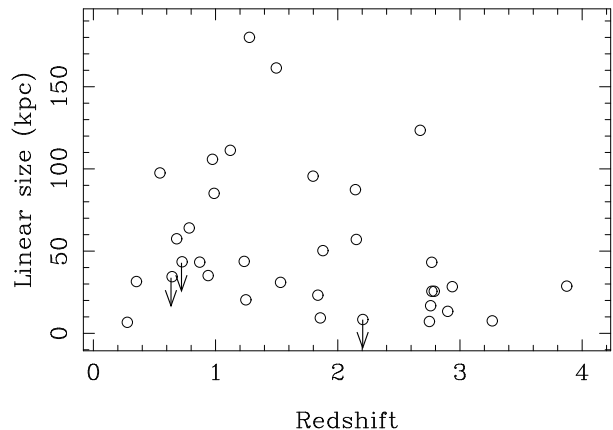,width=8cm}}
\caption{Largest linear size from Table~\ref{data} 
versus redshift for sources (excluding QSOs) in 
our sample with spectroscopic
redshifts. 
}
\label{LLS_z}
\end{figure}

In some USS-selected HzRG samples an explicit maximum radio angular size
cut-off has been applied \citep[see for example][]{Blund98,deB00a},
on the assumption that the higher-redshift sources 
are younger and hence smaller \citep{Blund99}.
Our plot of largest linear size
versus redshift in Fig.~\ref{LLS_z} is limited by the small number of redshifts
measured so far (21 per cent of $K$ detections have redshifts).
The selection of galaxies for redshift followup is likely to exclude some 
radio-extended but
optically-faint objects (as discussed in Section~\ref{Kaaz}) which we expect to be 
lower redshift dust-obscured galaxies that would
populate the top left of Fig.~\ref{LLS_z}.
However, the upper envelope of source sizes drops at higher redshift.
Linear size evolution is well known in radio samples \citep{Kap87, Bar88},
but \citet{Blu02} found only a weak correlation between redshift
and linear size measured at low (74\,MHz) frequency. The correlation here
is also weak.
Further analysis will be carried out when 
the spectroscopy for the MRCR--SUMSS sample is complete.

\begin{figure}
\centerline{\psfig{file=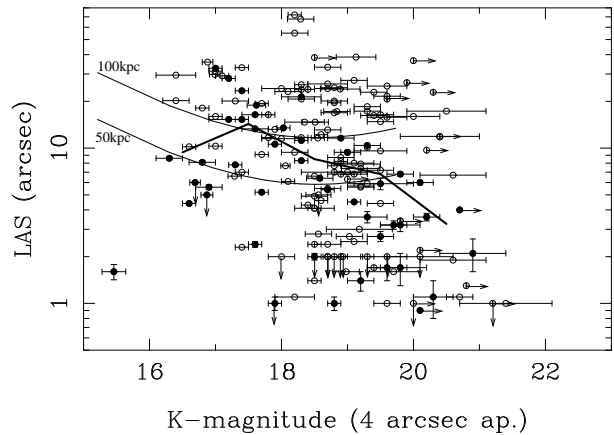,width=8.0cm}}
\caption{Largest angular size (LAS) 
versus 4-arcsec-aperture $K$ magnitudes for the MRCR--SUMSS sample. 
The LAS was measured from the 4800- or 8640-MHz images \citepalias[see][table 3]{Bry08} 
where
available, and otherwise, the value comes from \citetalias[][table 2]{Bry08}.
Two solid curves represent the
angular size of a 50-kpc- and 100-kpc-diameter galaxy versus redshift (based on
Equation 1 from the $K$--$z$ plot) for our
adopted cosmological model.
The heavier solid line joins points which are the median LAS values in each
magnitude bin from 16--17 up to 20--21 magnitudes. Solid points represent the objects
that were followed up with spectroscopy (irrespective of whether that resulted in a redshift
measurement). 
}
\label{LAS_K}
\end{figure}

The measured $K-$magnitudes have also been plotted
against largest angular size (LAS) in Fig.~\ref{LAS_K}. 
Our adopted cosmology dictates that the LAS will
change with redshift, and therefore a line is shown 
representing the LAS of a 50-kpc source at each redshift, and similarly for
a 100-kpc source. In both cases, the LAS was calculated using our
chosen cosmological parameters (see Section~\ref{Intro}), and then the
redshifts were converted to $K$ magnitudes using Equation~\ref{eq1}.
If the distribution of points decreases faster than the 
lines marking the cosmological models, then more distant 
USS-selected galaxies may be intrinsically smaller than their low-redshift
counterparts. 
The median LAS of our objects at each $K$ magnitude generally
decreases faster than the cosmological models.
However, a Spearman rank test confirms there is no correlation
in Fig.~\ref{LAS_K}.

\subsection{The alignment effect}

\begin{figure}
\centerline{\psfig{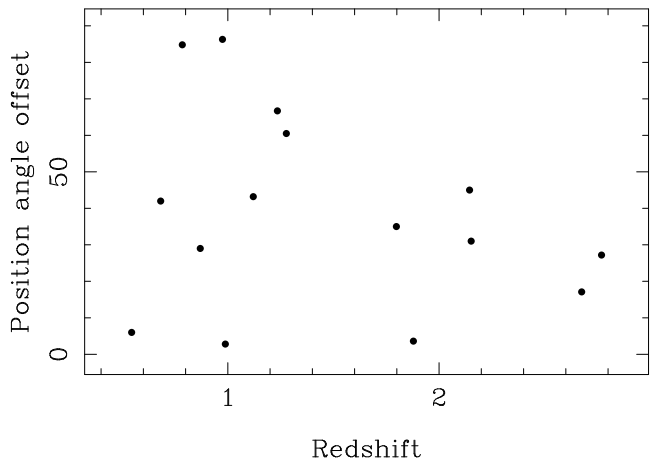}}
\caption{The offset between the radio axis position angle
and that of the major axis of the ellipsoidal fit to the $K$-band galaxy versus redshift.
Sources with LAS$<$5\,arcsec are not included as the position angle offsets may
be less accurate.
}
\label{PAvsz}
\end{figure}

Alignment between near-infrared and radio structures has been found in
many HzRG samples, and these are discussed in detail in \citetalias{Bry08}.  
However, the redshift dependence of the alignment effect is very much in
dispute, with some investigations finding a clear increase in alignments
with redshift \citep[e.g.][]{vanB98}, 
while others find no such
trend \citep[e.g.][]{Pen99}. 
We now examine the redshift
dependence of the alignment effect in the MRCR--SUMSS sample.  In \citetalias{Bry08}
we fitted ellipsoidal profiles to the $K$-band host galaxies. The major
axis of the ellipsoids was found to align with the radio structure in 21
per cent of the 122 galaxies for which an elliptical profile could be
fitted. Fifteen of the 122 fitted galaxies have redshifts and they are
plotted in Fig.~\ref{PAvsz} against the position angle offset, defined as
the difference between the radio axis and the major axis of the fitted
$K$-band ellipsoid. 
There may be an
upper envelope to the distribution of points, which falls with 
increasing redshift. If that remains the case 
when the spectroscopic sample is complete, then 
it would imply that the higher-redshift objects 
show closer alignment than those at lower redshifts.

In \citetalias{Bry08} we identified sources with alignment offsets of less than 10
degrees as being likely candidates for jet-induced star formation. Only
three of those have redshifts in Fig.~\ref{PAvsz}. While the trend with redshift
identified here is clearly not associated with such highly aligned
processes, larger position angle offsets have been attributed to
ionisation cones or extended emission-line regions in some high redshift 
galaxies. Such features would not be
seen in galaxies at $1.3\leq z\leq 1.9$ where there are no bright emission 
lines in the $K$-band.
It is noteworthy that
the objects which are highly misaligned, including those with
perpendicular axes ($>80$\,degrees), are all at low redshift. 

\subsection{The efficiency of low-frequency selection}

\subsubsection{Spectral curvature}
The original purpose of selecting sources that are ultra-steep between 408--843\,MHz
was to test whether the sample would preferentially select higher redshifts than USS samples
selected at higher frequencies. This would be the case if the radio spectrum
steepens at higher frequencies and the steeper part is redshifted down to
408--843\,MHz, in which case the lower-frequency selection would include fewer  
low-redshift objects.
In
\citetalias{Bro07} we found that 85 per cent of USS-selected sources have straight radio spectra
between 408--2368\,MHz (observed frame) and therefore
the frequency of USS selection should not affect the resultant sample
if selected at a frequency higher than 408\,MHz. 
Similarly \citet{kla06} found that 89 per cent
of a subsample of 37 sources from the SUMSS--NVSS sample also have 
straight radio spectra from 843\,MHz to either 6.2, 8.6 or 18\,GHz. However, 
\citet{bor07} found that the majority of their USS sample
have radio spectra that flatten from 352\,MHz to 74\,MHz. If the radio spectrum remains
straight over a larger frequency range, then the spectral index is independent of the
observed frequency. In that case, the $z$--$\alpha$ correlation would not be due to
a k-correction. The steep
spectrum would instead be intrinsic to the source or the environment it is in. 

We have an opportunity to investigate the shape of the radio spectrum below 408\,MHz for the
MRCR--SUMSS sample as some of the sources are also in the
74-MHz VLA Low-Frequency Sky Survey \citep[VLSS;][]{Coh:07} 
and the 151-MHz Mauritius Radio Telescope (MRT) Southern Sky Survey \citep{Pan:06}.
These two catalogues, while useful, have limitations. Firstly, the 
declination of our sources is at the southern edge of the VLSS survey region,
where the flux densities are considered less reliable. Secondly, the 
large MRT beamwidth ($4\times4.8$\,arcmin at declination $-35^{\circ}$) is
subject to beam confusion. We note that the MRT flux density calibration
is tied to interpolated 151-MHz flux densities for strong sources in the
\citet{bur06} 
compilation. Errors in flux density were
assumed to be the quadratic combination of a 0.26-Jy-rms noise plus
confusion term and a 6.3 per cent scale uncertainty \citep{Pan:06}.

\begin{figure}
\centerline{\psfig{file=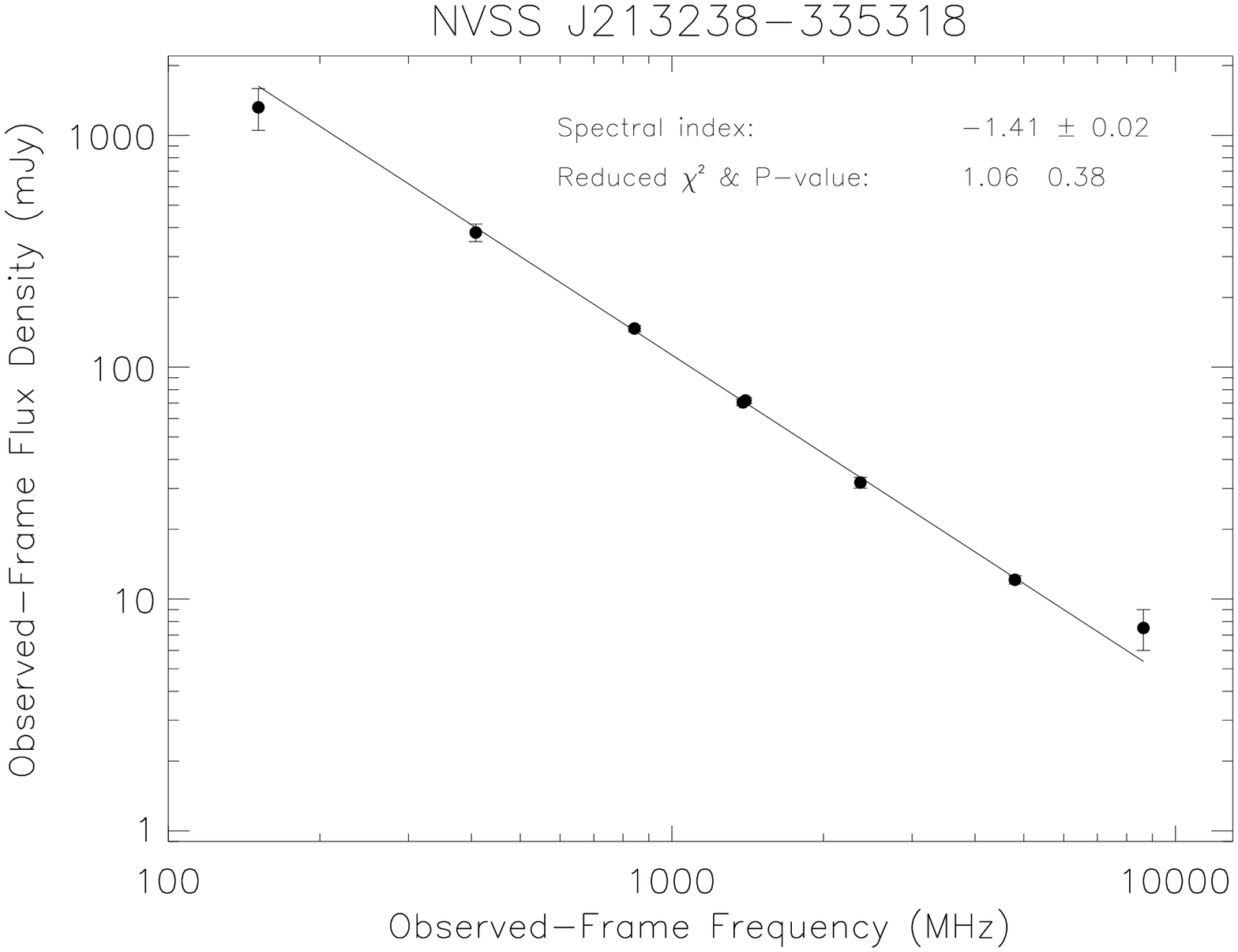,width=6.5cm,angle=90}}
\caption{The observed-frame radio spectral energy distribution for 
NVSS~J213238$-$335318 at $z=2.900$. Of the sources in Tables~\ref{151} and ~\ref{74},
this source was fitted to a power-law using the largest frequency range, from 
151 to 8640\,MHz. The resultant spectral index and goodness-of-fit statistics
are shown.}
\label{straightspec}
\end{figure}

Inclusion of 74- and 151-MHz flux densities in our spectrum fits was
done only after careful assessment of the radio images. MRT  
and VLSS images were overlaid with
images from MRCR and SUMSS to check for confusion in the MRT images,
and flux density or morphological anomalies in VLSS at these large zenith
distances.
The MRCR--SUMSS
sample sources which have a reliable 74-MHz flux density have been fitted with a 6-point
spectral index between 74 and 2368\,MHz in \citetalias{Bro07} (table\,10) and those with
a $K$-band counterpart are listed in Table~\ref{74}. 
The 151-MHz catalogue does not cover all of our
survey region, but we found matches to ten of the MRCR--SUMSS sources which have
$K$-band counterparts. These are given in
Table~\ref{151} along with a revised spectrum fit which includes the new frequency point.
Two of these have both 74-MHz and 151-MHz flux densities.
An example of the fit to one source is shown in Fig.~\ref{straightspec}.

\begin{table*}
\caption{Sources from the MRCR--SUMSS sample which have a $K$-band identification
and a 151-MHz flux density
in the MRT catalogue. $K$ magnitudes in a 4-arcsec aperture and the 151-MHz
flux densities are listed. 
The spectral index is shown for sources with 
a power-law radio spectrum. All the spectral indices are 6-point values between 151 and 2368\,MHz,
except for NVSS~J213238$-$335318 which has an 8-point index from 151 to 8640\,MHz, NVSS~J233729$-$355529 
which has a 4-point index from 151 to 1400\,MHz and NVSS~J234145$-$350624 which has fluxes measured at 
9 frequencies from 151\,MHz to 18\,GHz. 
Sources without spectral indices listed flatten towards lower frequencies. 
Objects which also
have 74-MHz flux densities \citepalias[][table 10]{Bro07} are marked with an asterisk, and the 74-MHz
flux density was used in the spectral index fit.}
\begin{tabular}{lcccc}
\hline
\hline
Source                 & $K$ & S$_{151}$ & $\alpha$ & Redshift\\
                       & (4-arcsec) & (Jy) &  & \\
\hline
NVSS~J003445$-$372348  & 17.0  & $2.77\pm0.31$ & $-1.20 \pm 0.03$ & --- \\
NVSS~J202720$-$341150  &  $>20.3$  & $2.39\pm0.30$ & $-1.13 \pm 0.04$ & --- \\
NVSS~J213238$-$335318  &  19.8 & $1.32\pm0.27$ & $-1.41 \pm 0.02$ & 2.900\\
NVSS~J214114$-$332307$^{*}$  & 19.1  & $2.19\pm0.29$ & flattening & --- \\
NVSS~J215455$-$363006  & 17.7  & $3.78\pm0.35$ & $-1.13 \pm 0.03$ & 1.235\\
NVSS~J223305$-$365658  & 16.6  & $1.99\pm0.29$ & $-0.95 \pm 0.04$ & 0.939\\
NVSS~J233535$-$343330  & 18.9  & $3.40\pm0.34$  & $ -1.05 \pm 0.03$ & ---\\
NVSS~J233729$-$355529  & 19.2  & $2.58\pm0.31$  & $ -1.39 \pm 0.05$ & --- \\
NVSS~J234145$-$350624  & 16.9 & $11.26\pm0.76$ & flattening & 0.644 \\  
NVSS~J235945$-$330354$^{*}$  & 19.1 & $2.76\pm0.31$  & flattening & ---\\
\hline
\label{151}
\end{tabular}
\end{table*}
\begin{table*}
\begin{minipage}{175mm}
\caption{Sources from the MRCR--SUMSS sample which have 
a 74-MHz flux density \citepalias[from][table 10]{Bro07} 
and a $K$-band identification and are not already shown in Table~\ref{151}. $K$ magnitudes in a 4-arcsec aperture and the 74-MHz
flux densities are listed. 
All the spectral indices are six-point values between 74 and 2368\,MHz except for NVSS~J230846$-$334810 
which has a four-point fit between 74 and 1400\,MHz.
Sources without spectral indices listed flatten towards lower frequencies. 
}
\begin{center}
\begin{tabular}{lcccc}
\hline
\hline
Source                 & $K$ & S$_{74}$ & $\alpha$ & Redshift\\
                       & (4-arcsec) & (Jy) &  & \\
\hline
NVSS~J000231$-$342614  & 18.2  & $1.35\pm0.19$ & $-0.98 \pm 0.04$ & --- \\
NVSS~J000742$-$304325  & 18.9  & $0.77\pm0.10$ & flattening & --- \\
NVSS~J001506$-$330155  & 17.6  & $2.10\pm0.24$ & $-0.89 \pm 0.03$ & --- \\
NVSS~J021759$-$301512  & 18.3  & $1.56\pm0.17$ & $-1.04 \pm 0.03$ & --- \\
NVSS~J120839$-$340307  & 18.0  & $22.93\pm2.34$ & $-1.10 \pm 0.03$ & 1.120 \\
NVSS~J141428$-$320637  & 15.5\rlap{$^{\it a}$}  & $3.78\pm0.49$ & flattening & 0.2775 \\
NVSS~J215047$-$343616  & 17.4  & $4.49\pm0.53$ & $-1.05 \pm 0.03$ & --- \\
NVSS~J215226$-$341606  & 18.3  & $1.67\pm0.22$ & $-0.99 \pm 0.03$ & 1.277 \\
NVSS~J221650$-$341008  & 16.4  & $2.23\pm0.29$ & $-1.00 \pm 0.03$ & --- \\
NVSS~J230846$-$334810  & 17.0  & $1.61\pm0.20$ & $-1.13 \pm 0.04$ & --- \\
\hline
\label{74}
\end{tabular}
\vspace{-1cm}
\footnotetext[1]{2MASS magnitude in an 8-arcsec aperture.}
\end{center}
\end{minipage}
\end{table*}

In Tables~\ref{151} and ~\ref{74}, 75\,per cent (15/20) of sources have
spectral shapes that remain a power-law down to either 74 or 151\,MHz. 
The mean
$K$-band magnitudes are 17.9 and 18.0 for the flattening and straight-spectrum sources
respectively.
The difference
between these distributions is not statistically significant (based on a Wilcoxon rank-sum test).
Therefore, we find that the majority of our sources have a spectral shape that remains
straight, which supports a picture in which the steep spectrum is intrinsic to high redshift
sources and that a k-correction is not the reason why these high-redshift sources have been
identified by our selection criteria.

\subsubsection{Relationships among redshift, $K$, $\alpha$ and 1400\,MHz flux density distributions}
\label{Kaaz}
In \citetalias{Bry08} we compared the $K$-band magnitude distribution of
the MRCR--SUMSS sample to both USS-selected- and non-USS-selected-samples
from the literature, and found that the resulting $K$-band distribution was
independent of the flux density distribution. Here we look at the redshift
distribution of the same samples and 
we investigate any biases in the selection of objects
for spectroscopic followup.

\begin{figure*}
\psfig{file=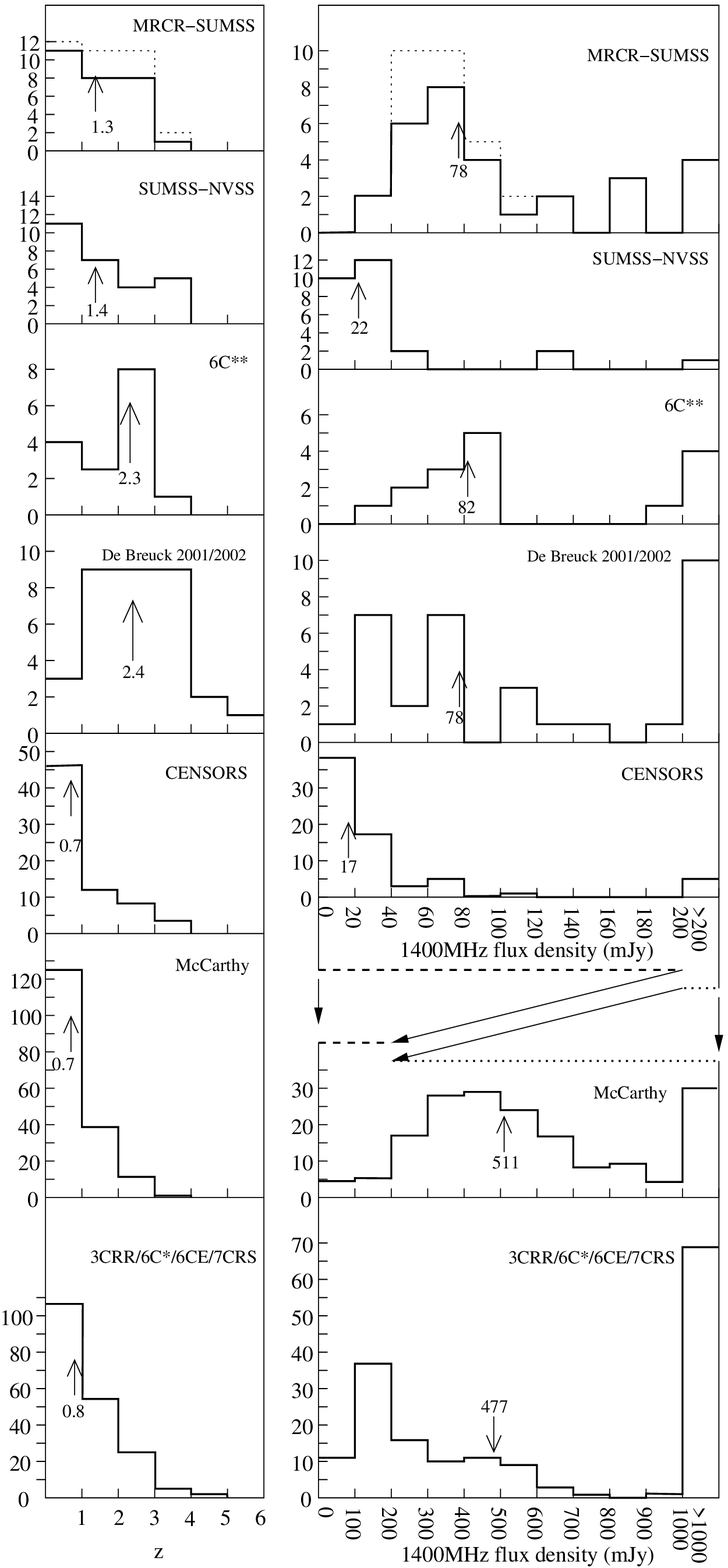, width=8.5cm}
\caption{Distributions of redshifts and 1400\,MHz flux densities for the
samples from MRCR--SUMSS, SUMSS--NVSS \citep{deB04,deB06},
\citet{deB01,deB02}, 6C** \citep{cru06}, 
CENSORS \citep{bro06,bro08},
\citetalias{McC96} 
and a combined sample from the 3CRR, 6C$^{*}$, 6CE and 7CRS catalogues \citep[compiled by][]{Wil03}. \citetalias{Bry08} included histograms of the flux 
densities of all the sources with $K$-band magnitudes, while here we show the
flux densities of the subset of sources with redshifts.
The CENSORS redshift histogram includes both the galaxies
for which we had calculated a 4-arcsec magnitude in \citetalias{Bry08} as well
as those that only had 3-arcsec-aperture magnitudes and hence are not in the
$K$-magnitude histogram of \citetalias{Bry08}.
Both the McCarthy and \citet{deB01,deB02} samples only have data
shown if the $K$ magnitude was measured, while sources with a redshift
but no $K$ magnitude are not included. This was done because we are comparing 
the redshift distributions of $K$-band-selected galaxies and those which did
not have $K$ magnitudes were based on different selection criteria.
Vertical arrows mark the median values in each
plot. QSOs have been removed from the literature samples. The redshift 
distribution for the MRCR--SUMSS sample shows both the
radio galaxies and QSOs (dotted line) and just the radio galaxies (solid line), but the median
calculation did not include the QSOs, in order to be directly comparable, to
the literature samples.
}
\label{hists}
\end{figure*}

Fig.~\ref{hists} shows the redshift and 1400\,MHz flux density
distributions for our MRCR--SUMSS sample and the SUMSS--NVSS, 
6C** \citep{cru06}, \citet{deB01,deB02}, 
\citetalias{McC96}, 
CENSORS \citep{bro06,bro08} and a combined sample from the 3CRR, 6C$^{*}$, 6CE and 7CRS catalogues \citep[compiled by][]{Wil03}. 
The McCarthy and CENSORS samples were not USS-selected, 
while the Willott et al. sample is predominantly non-USS-selected apart from 
23 sources from the 6C*. 
The McCarthy sample includes 175 redshifts from a set of 277 
$K$-band observations of radio galaxies. It is the largest non-USS $K$-band sample 
even though it remains unpublished. The 1400\,MHz flux density values
for the McCarthy sample were obtained by cross-matching the positions with the NVSS catalogue
within a 60-arcsec search radius. For sources which had
two matches within 60\,arcsec, we selected the closest match.
SUMSS--NVSS was selected to have $\alpha_{843}^{1400}<-1.3$, and 
the 6C** sample is from the 
151-MHz 6C survey with $\alpha_{151}^{1400}<-1.0$. 
The \citet{deB01,deB02} sample was selected at one of the
frequencies 325, 365, or 408\,MHz as detailed in the papers,
and has $\alpha_{\sim350}^{1400} \lesssim -1.3$. We have only included sources from
the \citet{deB01,deB02} sample that have measured $K$ magnitudes.
The combined sample compiled by \citet{Wil03} includes 202 narrow-line galaxies
from the 3CRR, 6C$^{*}$, 6CE and 7CRS catalogues, selected at 151 and 178\,MHz.
Spectroscopic redshifts are available for 193 ($96$ per cent) of those.
The 1400\,MHz flux densities were
obtained from the NASA/IPAC Extragalactic Database (NED) by selecting the most
recently published\footnote{Flux densities came from \citet{Con98}, \citet{Lai80}, \citet{Kel69}, \citet{Whi92}, \citet{Owe97}, \citet{Cro05} and \citet{Bec95}.} integrated flux density. In several cases only peak
flux densities were available and 25 targets ($13$ per cent) have no published observations at
1400\,MHz. As the 3CRR/6C$^{*}$/6CE/7CRS combined sample has 
nearly-complete redshift followup, the histograms shown in Fig.~\ref{hists} are
not biased compared to the corresponding $K$-band distributions in \citetalias{Bry08}. 

It is important to note that while the $K$-magnitude
measurements are mostly complete for each of the USS samples, the redshifts
are incomplete. 
The percentage of $K$-band detections with spectroscopic redshifts is
$32$ per cent for 6C**, $46$ per cent for SUMSS--NVSS, $40$ per cent for 
\citet{deB01,deB02} and currently $21$ per cent for
the present MRCR--SUMSS sample. 

The choice of targets for follow-up spectroscopy in the MRCR--SUMSS sample
was essentially determined by the telescope being used. On the one hand, the faintest
objects are the best high-$z$ candidates for followup spectroscopy, but the available telescope
time will always limit the number of faint objects that can be observed. The brightest $K$-band
objects were observed with the ANU 2.3\,m, and those that were not detected
were then observed on the NTT along with fainter targets down to $K \lesssim 19$.
After the magnitude cut, the targets were then chosen on Right Ascensions
available.
The $K>19$ objects were observed at the VLT, where we had a slight preference to
observe the more compact objects, but this was also limited by the accessible Right
Ascensions. 
To test whether the objects that have spectroscopy are representative of
the whole sample, we distinguish them with different symbols in Fig.~\ref{LAS_K} which shows
largest angular size of the sources versus $K$-band magnitude.
While the distribution of the objects with spectra is similar
to that for the
full sample, among the faintest targets observed on the VLT there is an
excess of the most compact sources.
Larger sources could be faint because they are genuinely distant or because they are closer 
but dust obscured.
The effect of
this selection on our redshift distribution would be a small bias towards higher-redshift 
objects. Nevertheless, the 
median $K$-band magnitude for the spectroscopic subset is 
18.3 compared with 18.7 for the
full sample (with known QSOs removed).  This indicates that the
spectroscopic sample is representative of the full sample, with the
slightly brighter median simply due to the fact that we have had more NTT
than VLT observing time. The USS-selected literature samples also have
median $K$-band magnitudes for their spectroscopic samples that are
$\leq0.2$ magnitudes brighter than for the full $K$-band samples.

A more stringent test of whether the spectroscopic samples are typical of the complete
samples is to compare flux density distributions.
In
Fig.~\ref{hists} we compare the flux density distributions from \citetalias{Bry08}
(fig.\,10) with those from the literature.
The sources with redshifts from the MRCR--SUMSS, 
SUMSS-NVSS, 6C**, CENSORS and McCarthy samples are all
consistent with being drawn randomly from the flux density distributions
of the whole $K$-band samples, based on Kolmogorov-Smirnov test $p$ values
of 0.97, 0.986, 0.910, 0.864 and 0.58 respectively.
The $p$ value of 0.099 for the \citet{deB01,deB02} sample indicates that the 
sources with redshifts from this sample are somewhat less representative of the
original $K$-band population.

Now that it is clear there are no strong biases in the $K$-band selection or the flux densities
of the sources that have redshifts, 
we can investigate
the corresponding redshift distributions.
The CENSORS, McCarthy and 3CRR/6C$^{*}$/6CE/7CRS redshift distributions in Fig.~\ref{hists} are 
weighted towards low redshifts, while the USS-selected samples
have returned higher median redshifts, as expected.
The different selection frequencies and spectral indices result in the
variation in flux densities between samples.
In \citetalias{Bry08} we found that the $K$-magnitude distribution did not vary as a
function of flux density.
The redshift distributions for
the USS samples also do not correlate with flux density.
On the one hand, the SUMSS--NVSS and MRCR--SUMSS samples have similar median redshifts
but markedly different flux densities. On the other hand, the two samples
netting the highest redshifts have high median flux densities. 

In \citetalias{Bry08}, we discussed how the k-correction would predict that
samples selected at lower radio frequencies should result in a higher median redshift.
Considering just the groups that have been shown to have redshifts
representative of the full sample, the MRCR--SUMSS, SUMSS--NVSS and 6C** 
have selection frequencies of 408, 843 and 151\,MHz 
respectively. For these three samples the median redshift does 
not correlate with selection frequency, supporting a picture in which the k-correction
is not responsible for the steep spectral-energy distribution of high-redshift sources.

An alternative explanation for the steep-spectral-energy distribution of radio galaxies 
was given by \citet{Ath98}. They suggested that 
higher-density regions at high redshift
reduce the hotspot velocities through the intergalactic medium, resulting in steeper
electron-energy distributions. \citet{kla06} 
then proposed that the mounting evidence
that HzRGs live in over-dense regions or protoclusters \citep[e.g.][]{Car97,Kur00b,Pen00,Ste05,Kod07,Ven:07} 
and the fact that their radio spectral
energy distributions are straight, implies that the $z$--$\alpha$ correlation is due to the
dense environment at high redshift.
Fig.~\ref{z-alpha} shows spectral index versus redshift for MRCR--SUMSS
and for three USS-selected samples and one non-USS-selected sample from the literature. All three 
USS-selected literature
samples have a nominal spectral index cut-off of $\alpha=-1.3$, while the MRCR--SUMSS 
cut-off was $-1.0$. The non-USS-selected sample is based on the data in \citetalias{McC96}, 
from which the sources with redshifts were cross-matched with the NVSS catalogue and a spectral index
was calculated between 408 and 1400\,MHz. 
Fig.~\ref{z-alpha} shows an apparent trend in spectral index
with redshift for the USS-selected sources which is not real; it arises
because the spectral index cut-off in each sample has been breached by
objects whose spectral index changed after sample selection because of
revised flux densities.
Neither the points in the literature samples with $\alpha<-1.3$, nor the
points in the MRCR--SUMSS sample with $\alpha<-1.0$ show a significant
correlation with redshift. 
In Fig.~\ref{hists}, the more stringent spectral index cut-off of
$\alpha<-1.3$ of the
SUMSS--NVSS sample did not result in a significantly higher redshift than did
the $\alpha<-1.0$ selection of the MRCR--SUMSS and 6C** samples. 
However, Fig.~\ref{z-alpha} clearly illustrates that a spectral index cut-off of $\alpha=-1$ eliminates
the bulk of the low redshift ($z<1$) sources. Applying a steeper 
spectral index cut-off will miss many $z>2$ objects. For the
objects shown, $\sim54$\,per cent of the $z>2$ galaxies, or $\sim49$\,per cent of the $z>2.5$ galaxies, 
would be missed by an $\alpha=-1.3$ cut-off. 
However, the $z$--$\alpha$ correlation is most clearly defined by the lower envelope in the 
$z$--$\alpha$ plane, which is seen best in the non-USS-selected points.

If the higher-density environment at high redshift drives the steep radio spectra, then it is
interesting to note that in Fig.~\ref{z-alpha}, the sources steeper than $\alpha<-1.3$ are
no more likely to be at high redshift than at low redshift. However, the source numbers are low,
so we have also plotted the full MRCR--SUMSS $K$-band sample against spectral index in  
Fig.~\ref{K-alpha}.
While an individual $K$ magnitude is not a direct indicator of
redshift (see Section~\ref{Kztrends}), if the net distribution of $K$
magnitudes becomes fainter at higher redshift then the $\alpha$ versus $K$ plot in 
Fig.~\ref{K-alpha} should show a trend for sources with
$\alpha<-1.0$. However, there is no evidence that the USS sources 
steepen further with fainter $K$ magnitudes, suggesting there is no
advantage in selecting a more stringent spectral index cut-off.

\begin{figure}
\centerline{\psfig{file=Fig10.ps,width=6.0cm, angle=-90}}
\caption{Spectral index versus spectroscopic redshift for our MRCR--SUMSS sample (circles; using
$\alpha_{7{\rm -point}}$ from \citetalias{Bry08} table\,3 or, for sources without 4800- and 
8640-MHz radio data, we used
$\alpha_{5{\rm -point}}$ or $\alpha_{843}^{1400}$ from \citetalias{Bry08} table\,2)
and three samples from the literature (stars): \citet{deB01,deB02} (using 
$\alpha_{\sim350}^{1400}$), SUMSS--NVSS sample (using linear fitted spectral indices from 
\citet{kla06}, or when unavailable, $\alpha_{843}^{1400}$) and
\citet{bor07} ($\alpha_{325}^{1400}$). The dotted line shows
the spectral index cut-off of the literature samples (except a small number of sources
from \citet{deB01,deB02} had a cut-off at $\alpha<-1.2$), while our MRCR--SUMSS cut-off
is at $\alpha=-1.0$. In each sample there are some sources which have 
shallower spectral indices than the cut-offs because of subsequent revisions of source 
flux densities in the catalogues. The small dots are from the non-USS-selected sample
of \citetalias{McC96},  
with a spectral index measured between 408 and 1400\,MHz (see text for details).}
\label{z-alpha}
\end{figure}

\begin{figure}
\centerline{\psfig{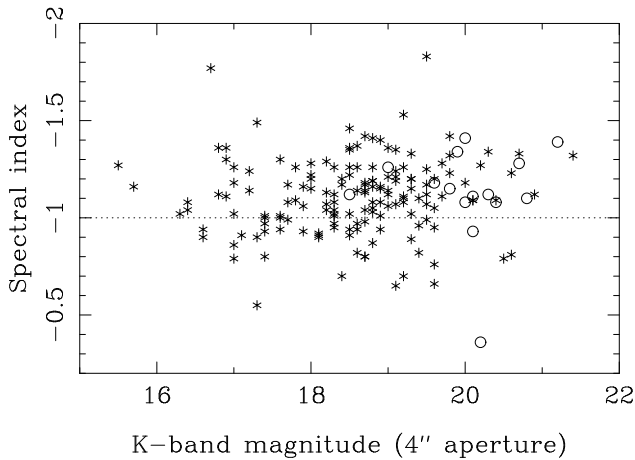}}
\caption{Spectral index versus $K$-band magnitude in a 4-arcsec aperture for our 
MRCR--SUMSS sample. The spectral indices are the 7-point fitted values from
\citetalias{Bry08} table\,3 for sources that have 4800- and 8640-MHz radio data, 
otherwise we used the 5-point fitted values 
or, when unavailable, the $\alpha^{1400}_{843}$
values as defined in \citetalias{Bry08} (table\,2). Our spectral index
cut-off was defined using 408--843\,MHz, and therefore some of our sources
turned out to be flatter when fitted over a large frequency range, and 
some had the 843-MHz flux densities revised by the new SUMSS catalogue (version 2.0), as discussed
in \citetalias{Bro07}. The sources with $K$-band magnitude limits are marked by a circle rather than a
right-pointing arrow, for clarity. For sources steeper than $-1.0$, there is no correlation between
spectral index and $K$ magnitude.
}
\label{K-alpha}
\end{figure}

So far, USS-selected samples have been
efficient at finding high-redshift radio galaxies, although non-USS samples have
also identified many $z>2$ objects that may well have been missed by
USS selection \citep[e.g.][]{bro06,bro08}.
While the USS-selection nets higher-redshift samples (Fig.~\ref{hists}), a secondary
gain is that USS-selection requires substantially less observing to return the
same redshift benefits. 
For example, CENSORS has 122 sources in a 6 deg$^{2}$ 
field, while the USS-selected SUMSS--NVSS and MRCR--SUMSS samples have 65 and 162 (160 with
4-arcsec-aperture magnitudes)
$K$-band objects respectively across the sky. Considering only the $K\geq19$
galaxies which are potentially the most distant,
the USS-selected samples have $>1.5$ times the percentage of
$K\geq19$ objects than the non-USS sample.
Of the $K\geq19$
sources with spectroscopic redshifts, the USS-selected samples have
a larger fraction of high-redshift ($z>2.5$) objects, but
the numbers are small, especially for the SUMSS--NVSS sample.
This means that 
the non-USS sample would require at least 1.5 times more $K-$band imaging than
the USS samples
in order to net a smaller percentage of high redshifts. Therefore, if the aim 
is principally to get a high-redshift sample of objects, then USS selection will
reduce the amount of imaging required.

\subsection{Spectral line diagnostics}
\label{Speclinediag}

\begin{figure*}
\centerline{\psfig{file=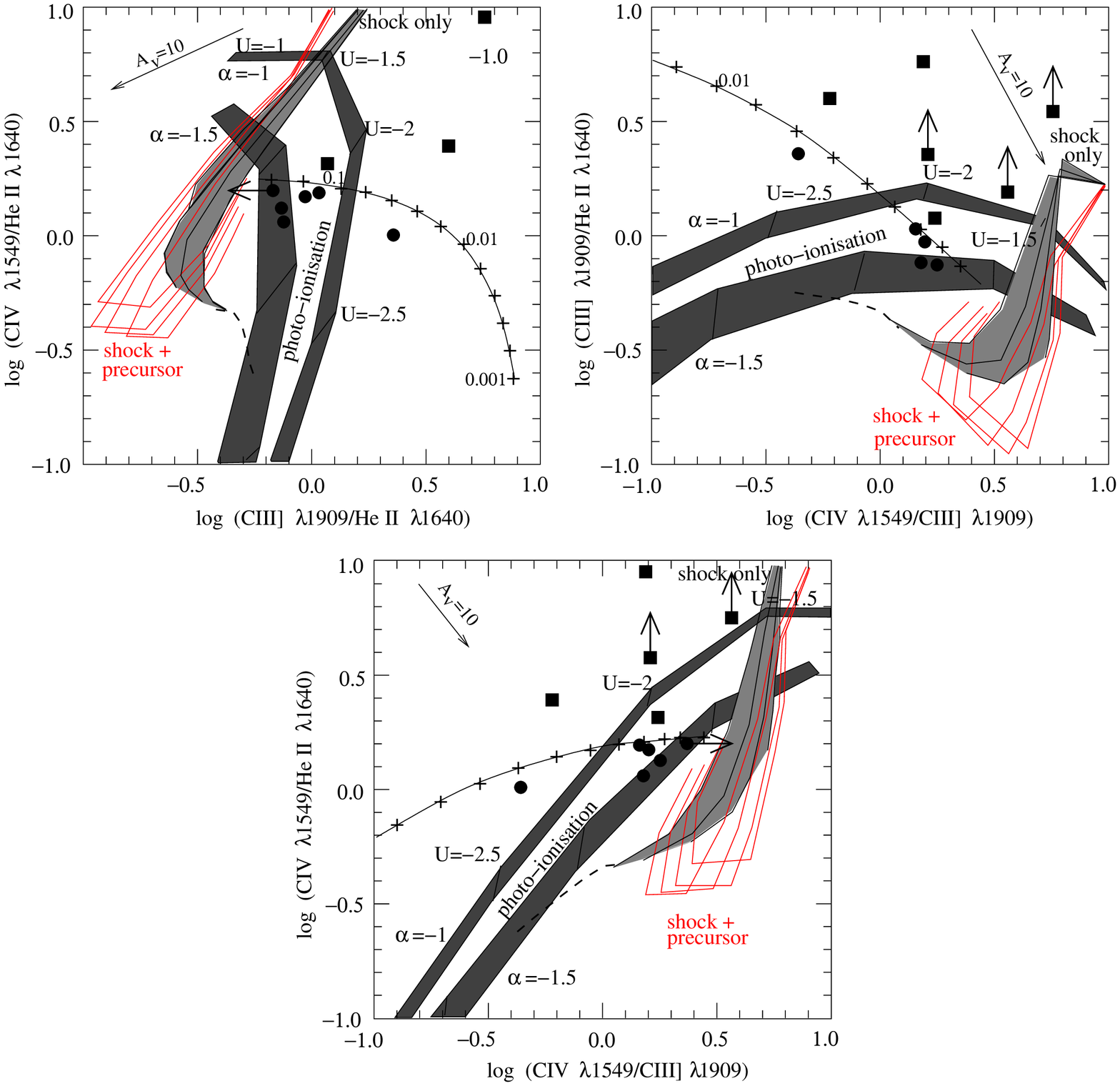, width=16cm}}
\caption{$UV$ spectral line diagnostic plots for the lines \civ, \ciii\ and \heii.
The galaxies in our sample that have at least two of these lines 
have been plotted (dots). In some cases an upper limit has been used for one of 
the lines. Radio galaxies are marked by dots, while the QSOs have square symbols.
We have also plotted photoionisation and shock models reproduced directly from 
\citet{deB00b}; the details of the model parameters are discussed in that paper. 
The light grey shaded grid represents shock models
while the unshaded grid (which partially overlaps the shock models) shows
the shock$+$precursor models. The shock velocities start at 150\,km\,s$^{-1}$ at the 
top right of the curves, and increase along the curves to 500\,km\,s$^{-1}$ 
at the end. The four curves in each grid represent four values of the magnetic 
parameter $0<B/\sqrt{n}<4 \mu$G\,cm$^{3/2}$. Shock models for 500--1000\,km\,s$^{-1}$ 
with $n=1.0$ and $B=3.23$ from \citet{All08} are marked by a dashed line. The dark grey shaded grids 
represent photoionisation models with four photo-ionisation sequences, with 
values of the ionisation parameter $\log_{10}(U)$ every 0.5 dex. The top and
bottom shaded photoionisation
grids have two models each, with power law spectral indices of 
$\alpha=-1$ and $\alpha=-1.5$ respectively. The left or bottom edge of the
photoionisation grids are models with hydrogen density $n=100$\,cm$^{-3}$ and the top or
right edges have $n=1000$\,cm$^{-3}$. The single line with crosses along it represents
the model of \citet{Bin96}. This model assumes that the photoionised light has gone 
through both matter-bound and ionisation-bound clouds in which the ratio
of the solid angle subtended by the matter-bounded to ionisation-bounded
clouds is varied from 0.01 to 100 along the line with tick marks every 
0.02\,dex. An arrow represents 10\,magnitudes of extinction based on the Milky
Way extinction law from \citet{All98}. Two QSOs (NVSS~J152123$-$375708 and
NVSS~J152435$-$352623) have upper limits on the flux of their \heii\ lines which 
give C$\,${\sc iv}/He$\,${\sc ii}$>1$  
and for NVSS~J152435$-$352623 C$\,${\sc iii}]/He$\,${\sc ii}$>1$ and therefore 
these values are outside the region plotted.
}
\label{diag}
\end{figure*}

The bright $UV$ emission lines in HzRGs can be produced by
one of several ionisation mechanisms.
The central AGN can photoionise the emission-line region gas,
while shock ionisation can result from the expanding cocoon associated with
the radio jets, which shocks the lower temperature gas clouds embedded
within the higher temperature gas surrounding the AGN.
In high-redshift galaxies, $UV$ emission lines are shifted 
into the optical band, and can be used to distinguish between 
shock and photoionisation
processes.
\citet{deB00b} discuss in detail the different models for
shock excitation \citep[based on the models of][]{Dop96} and photoionisation
(based on models by \citeauthor*{Vil97} \citeyear{Vil97} and 
\citeauthor*{All98} \citeyear{All98}) 
resulting in a set of diagnostic line-ratio
plots. In Fig.~\ref{diag}, we have reproduced those models on the plots 
of the line ratios for our galaxies. 
There are 13 spectra in 
Fig.~\ref{fig:spectra} which have at least two of the $UV$ diagnostic lines
\civ, \heii\ and \ciii.  
For the sources which show only two of the three lines, we 
determined $3\sigma$\ upper limits to the
flux from the missing line. Following \citet{Hob84}, 
the limiting flux
was calculated from the product of three times the rms fluctuation in the continuum
and a line width $\Delta v$.  For \ion{He}{ii} we set $\Delta v$
equal to the width of the \ion{C}{iv}\ line (since they are both
permitted lines with similar ionisation potentials), while for
\ion{C}{iii}] we set $\Delta v$\ equal to the instrumental resolution,
since this line is typically narrow in radio galaxies.
All of the galaxies plotted have $z>1.8$.

The diagnostic diagrams show that all  
of the galaxies are consistent with photoionisation as the primary excitation process.
There are no galaxies which can be explained by shocks alone or
shock plus precursor excitation, but 
there is one with a limiting arrow which may extend to the shock models.
The majority of our galaxies lie between the lines representing photoionisation
models where the incident photoionising continuum has a power law spectral
index of $\alpha>-1.5$ and ionisation parameters of $-2.5< \log_{10}(U)<-1.5$.
These results agree with the much larger sample of HzRGs presented in \citet{deB00b}.
\citet{McC93} 
produced a composite spectrum for $0.1<z<3$ radio galaxies, which also
demonstrated the dominance of photoionisation in radio galaxies.

The QSOs in Fig.~\ref{diag} typically lie above the galaxies in each plot due
to a comparatively smaller He$\,${\sc ii} line flux. The composite QSO 
spectrum from the Sloan Digital Sky Survey \citep{Vand01} has $\log$(C$\,${\sc iii}]/He$\,${\sc ii}$)=1.5$ 
and $\log$(C$\,${\sc iv}/He$\,${\sc ii}$)=1.7$, which are very much higher than 
the QSOs in our sample. The QSOs in our sample have ratios much closer to that
of the radio galaxies. Compact Steep-Spectrum (CSS) radio QSOs from \citet{Bak95} 
similarly showed stronger He$\,${\sc ii} than typical QSOs.

In \citetalias{Bry08}, we found that some sources in our sample show alignment
between the radio structures and either the major axis of the fitted $K$-band
ellipsoid, or the linear extended $K$-band emission. It was proposed that this
alignment could be due to jet-induced star formation or alternatively, scattering
of the nuclear emission along paths carved out by the jets through the gas. 
NVSS~J210814$-$350823, NVSS~J213637$-$340318 and NVSS~J223101$-$353227 have an ellipsoid major axis aligned within 10\,degrees of the radio
axis \citepalias[see][]{Bry08}, and are included on the emission line diagnostic plots in Fig.~\ref{diag}.
All three have line ratios that do not match the 
shock models. On the one hand, we may expect shock excitation in regions of jet-induced
star formation, but on the other hand, our spectra were extracted in $\sim1$-arcsec apertures,
which will be dominated by the nuclear source. Therefore we can not rule out jet-induced star
formation on the basis of the line ratios alone.

\begin{figure}
\centerline{\psfig{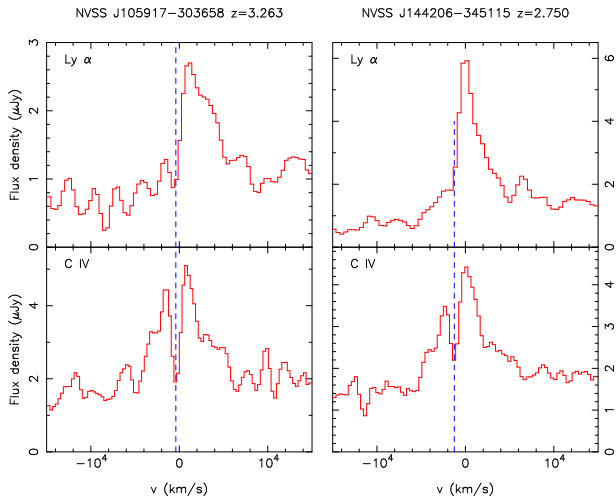}}
\caption{
Line profiles of Ly$\alpha$ and \civ\ on a common velocity scale for
the two sources which show self-absorption, NVSS~J105917$-$303658 and
NVSS~J144206$-$345115. The position of the common absorption is marked with
the dashed line. The velocity origin was determined by the redshift
in Table~\ref{data}.
}
\label{absorption}
\end{figure}

The spectra of HzRGs are typically dominated by the Ly$\alpha$ line in the
$UV$ spectrum. Ly$\alpha$ halos can extend to
hundreds of kiloparsecs around the radio source \citep{Kur00a,Ven:02,reu:07}. \hi\ gas surrounding the active
nucleus, or along the line of sight, can absorb Ly$\alpha$, cutting off the blue
wing of the spectral line. 
\citet{vanO97} and 
\citet[][fig. 11]{deB00b} 
have shown that
high-redshift galaxies with absorption in the Ly$\alpha$ line tend to have 
compact radio structures ($<50$\,kpc). While the resolution of our spectra
is insufficient to examine the Ly$\alpha$ profile in detail, there are two
quasars in the MRCR--SUMSS sample with obvious absorption in the blue wing of the Ly$\alpha$ line.
The absorption profiles are shown in Fig.~\ref{absorption}.
NVSS~J105917$-$303658 and NVSS~J144206$-$345115 have compact linear sizes of
7.6 and 7.2\,kpc, and have broad lines. 
The small apparent size may be due to alignment of the 
jets to the line-of-sight,
or the jets may be intrinsically compact. In the latter case,
these two
sources are consistent with the findings of \citet{vanO97} and 
\citet{deB00b} that stronger \hi\ absorption of the Ly$\alpha$ line may be due to a denser
medium surrounding and confining the radio source. 
An alternative model involves a low-density medium with an absorbing halo of low-metallicity
gas which has not yet been impacted by the radio jets \citep{Bin00}. 
Evidence for denser environments around these and other sources will be discussed further
in a subsequent paper based on Faraday rotation measures and galaxy
overdensities in the surrounding fields.

Dust in elliptical galaxies is typically associated with recent
mergers. If HzRGs are forming through hierarchical mergers, which may be the
trigger for the nuclear engine and hence the radio jets, then we might
expect the most compact, newly-triggered sources to show dust extinction. 
We now consider the evidence for dust from spectral lines.
While Ly$\alpha$ is generally the strongest $UV$ emission line
in HzRG spectra, it is not a good measure of dust as 
it is more likely to be be weak
relative to the other $UV$ lines as a result of viewing angle effects rather
than dust \citep{Vil96}. 
Based on the dust extinction vectors calculated by \citet[][fig.\,4]{All98}, 
the \ciii\ line is more affected by dust than \civ. Therefore, in spectra
such as that for NVSS~J215226$-$341606, the marginal detection of \ciii\ may be
indicative of dust obscuration, although there are insufficient other lines to
confirm this. The only other galaxy in which \ciii\ was not detected when 
\heii\ and/or \civ\ were detected is NVSS~J144932$-$385657. In this case,
based on an upper limit on the \ciii\ line, the diagnostic diagrams
from \citet[][fig.\,4]{All98} 
can explain the lack of \ciii\ with either
photoionisation or shock models, without the need for large dust extinction.

\section{CONCLUSIONS}

Spectroscopic results and analysis of the radio, $K$-band and redshift
data have been presented for a large Southern
hemisphere USS-selected HzRG survey, the MRCR--SUMSS sample. Our main
findings are:

(i) Based on
175 images and 52 spectra, 164 radio sources have $K$-band or SuperCOSMOS identifications 
with 36 confirmed galaxy redshifts. The highest new redshift found is 3.26, but
our spectroscopic followup is so far only 30 per cent complete. 
A comparison of
the USS samples with the non-USS-selected CENSORS and McCarthy samples shows that the
fraction of $z<1$ galaxies is significantly reduced by USS-selection, as expected.
However, for the USS-selected samples there is no correlation between the
median redshift and either median flux density or 
selection frequency.

(ii) Detailed analysis of the $K$--$z$ distribution includes a fit to
4-arcsec-aperture $K$ magnitudes for several samples from the literature.
This fit is compared with the 64-kpc fits of \citet{Wil03} and
\citet{deB04}. The MRCR--SUMSS sample agrees with the
\citet{Roc04} models for $>10^{11}$\mdot 
galaxies for all but two sources. We find no evidence for an increase in
dispersion in the $K$--$z$ plot at $z>2$ and conclude that these galaxies 
have undergone passive stellar evolution since $z>3$.

(iii) To test for curvature in the radio-spectral-energy distribution at 
frequencies lower than our selection frequency,
both the MRCR--SUMSS and the SUMSS--NVSS samples were cross-matched with
the 74-MHz VLSS catalogue and 151-MHz MRT catalogues. 
75\,per cent (15/20) of the sources with $K$-band magnitudes have radio-spectral-energy 
distributions that can be fitted by a single power-law.

(iv) The k-correction has frequently been assumed to account for the $z-\alpha$ correlation and
hence for the success of USS-selection in finding high-redshift-radio galaxies. The predominance of
straight spectrum objects over frequency ranges up to 74--8640\,MHz and the
lack of correlation with selection frequency are both evidence that an entirely different
mechanism is responsible for the steep spectra of high redshift radio galaxies.
Suggestions by \citet{kla06} that the steep spectrum is due to the higher-density
environments at high redshift will be tested for the MRCR--SUMSS sample in a subsequent paper.

(v) The spectral line ratios indicate 
that photoionisation is more dominant than shocks as the 
ionisation mechanism at the centre of these galaxies. QSOs in our sample 
have line ratios that are much closer to that of the radio galaxies than 
typical QSO ratios due to a stronger \heii\ line, which is also seen in CSS
sources. We find no evidence for significant dust extinction around the 
nucleus of any of our sources based on the line ratios.

Our method has been successful in identifying twelve $2<z<3.5$ galaxies
which are at the epoch when the star formation rate density of the Universe
was at a peak. They form an important sample for investigating
the environments and evolution of massive galaxies.
By the time the spectroscopic followup is complete, we expect more than 
40 galaxies in this redshift range.
In an upcoming paper,  we will present clustering and evolution studies based
on rotation measures and source overdensities in our images.

\bsp

\label{lastpage}

\subsection*{Acknowledgements}
We acknowledge financial support from the {\it Access to Major Research Facilities
Programme} which is a component of the {\it International Science Linkages Programme} established under the Australian Government's innovation
statement, {\it Backing Australia's Ability}.

Australian access to the Magellan Telescopes was supported through the Major National Research Facilities program of the Australian Federal Government.

JWB acknowledges the receipt of both an Australian Postgraduate Award and a 
Denison Merit Award. RWH, HMJ and JJB acknowledge support from the Australian 
Research Council and the University of Sydney Bridging Support Grants Scheme. 
BMG acknowledges the support of a Federation Fellowship from the Australian 
Research Council through grant FF0561298.

The expertise of Scott Croom was invaluable in the QSO identifications, and we really
appreciate his advice.
We thank Elaine Sadler for many useful discussions, Pat McCarthy for providing his unpublished data,
David Crawford
for providing the MRCR, and the team at the Magellan Telescopes for their exceptional efficiency and
good humour. We appreciate the assistance of the friendly staff at both the NTT and VLT.
We very much thank the referees for their detailed suggestions which improved the paper.

The Australia Telescope Compact Array is part of the 
Australia Telescope which is funded by the Commonwealth of Australia for 
operation as a National Facility managed by CSIRO. SuperCOSMOS Sky Survey 
material is based on photographic data originating from the UK, Palomar and 
ESO Schmidt telescopes and is provided by the Wide-Field Astronomy Unit, 
Institute for Astronomy, University of Edinburgh. This research has made use 
of the NASA/IPAC Extragalactic Database (NED) which is operated by the Jet 
Propulsion Laboratory, California Institute of Technology, under contract 
with the National Aeronautics and Space Administration.
IRAF is distributed by the National Optical Astronomy Observatories, which 
are operated by the Association of Universities for Research in 
Astronomy, Inc., under cooperative agreement with the National Science Foundation.
This publication makes use of data products from the Two Micron All Sky 
Survey, which is a joint project of the University of Massachusetts and the 
Infrared Processing and Analysis Center/California Institute of Technology, funded by the National Aeronautics and 
Space Administration and the National Science Foundation.


\begin{thebibliography}{}

\bibitem[Allen, Dopita \& Tsvetanov(1998)]{All98} Allen M. G., Dopita M. A., Tsvetanov Z. I., 1998, ApJ, 493, 571
\bibitem[Allen et al.(2008)]{All08} Allen M. G., Groves B. A., Dopita M. A., Sutherland R. S., Kewley L. J., 2008, ApJS, 178, 20
\bibitem[Antonucci(1993)]{ant93} Antonucci R., 1993, ARA\&A, 31, 473
\bibitem[Appenzeller et~al.(1998)]{aff+98} Appenzeller I. et al. 1998, ESO Messenger, 94, 1
\bibitem[Athreya \& Kapahi(1998)]{Ath98} Athreya R. M., Kapahi V. K., 1998, J. Astrophys. Astron., 19, 63
\bibitem[Baker \& Hunstead(1995)]{Bak95} Baker J.~C.,  Hunstead R.~W., 1995, ApJ, 452, L95
\bibitem[Baker et~al.(2002)]{bha+02} Baker J.~C.,  Hunstead R.~W.,  Athreya R.~M.,  Barthel P.~D.,  de Silva E., Lehnert M.~D.,    Saunders R. D.~E.,  2002, ApJ, 568, 592
\bibitem[Barthel \& Miley(1988)]{Bar88} Barthel P. D., Miley G. K., 1988, Nature, 333, 319
\bibitem[Barthel(1989)]{Bar89} Barthel P. D., 1989, ApJ, 336, 606
\bibitem[Becker, White \& Helfand(1995)]{Bec95} Becker R. H., White R. L., Helfand D. J., 1995, ApJ, 450, 559
\bibitem[Bhatnagar, Krishna \& Wisotzki(1998)]{bha98} Bhatnagar S., Krishna G., Wisotzki L., 1998, MNRAS, 299, L25
\bibitem[Bicknell et al.(2000)]{Bic00} Bicknell G. V., Sutherland R. S., van Breugel W. J. M., Dopita M. A., Dey A., Miley G. K., 2000, ApJ, 540, 678
\bibitem[Binette, Wilson \& Storchi-Bergmann(1996)]{Bin96} Binette L., Wilson A., Storchi-Bergmann T., 1996, A\&A, 312, 365
\bibitem[Binette, Kurk, Villar-Mart\'{i}n, R{\"o}ttgering(2000)]{Bin00} Binette L., Kurk J. D., Villar-Mart\'{i}n M., R{\"o}ttgering H. J. A., 2000, A\&A, 356, 23 
\bibitem[Blundell et al.(1998)]{Blund98} Blundell K. M., Rawlings S., Eales S. A., Taylor G. B., Bradley A. D., 1998, MNRAS, 295, 265
\bibitem[Blundell, Rawlings \& Willott(1999)]{Blund99} Blundell K. M., Rawlings S., Willott C. J., 1999, AJ, 117, 677 
\bibitem[Blundell et al.(2002)]{Blu02} Blundell K. M., Rawlings S., Willott C. J., Kassim N. E., Perley, R., 2002 NewAR, 46, 75.
\bibitem[Bock et al.(1999)]{Boc99} Bock D. C-J., Large M. I., Sadler E. M., 1999, AJ 117, 1578
\bibitem[Bornancini et al.(2007)]{bor07} Bornancini C. G., De Breuck C., de Vries W., Croft S., van Breugel W., R{\"o}ttgering H., Minniti D., 2007, MNRAS, 378, 551.
\bibitem[Bower et al.(2006)]{Bow:06}   Bower R. G., Benson A. J., Malbon R., Helly J. C., Frenk C. S., Baugh C. M., Cole S., Lacey C. G., 2006, MNRAS, 370, 645
\bibitem[Broderick et al.(2007)]{Bro07} Broderick J. W., Bryant J. J., Hunstead R. W., Sadler E. M., Murphy T., 2007 MNRAS, 381, 341 (Paper\,I)
\bibitem[Brookes et al.(2006)]{bro06} Brookes M. H., Best P. N., Rengelink R., R{\"o}ttgering H. J. A., 2006, MNRAS, 366, 1265.
\bibitem[Brookes et al.(2008)]{bro08} Brookes M. H., Best P. N., Peacock M. H., R{\"o}ttgering H. J. A., Dunlop J. S.,  2008, MNRAS, in press.
\bibitem[Bryant et al.(2009)]{Bry08} Bryant J. J., Broderick J. W., Johnston H. M.,Hunstead R. W., Gaensler B. M., De Breuck C., 2009, MNRAS submitted (Paper\,II)
\bibitem[Burgess \& Hunstead(2006)]{bur06} Burgess A. M., Hunstead R. W., 2006, AJ, 131, 100
\bibitem[Carilli et al.(1997)]{Car97} Carilli C. L., R{\"o}ttgering H. J. A., van Ojik R., Miley G. K., van Breugel W. J. M., 1997, ApJS, 109, 1
\bibitem[Cohen et al.(2007)]{Coh:07} Cohen A. S., Lane W. M., Cotton W. D., Kassim N. E., Lazio T. J. W., Perley R. A., Condon J. J., Erickson W. C., 2007, AJ, 134, 1245
\bibitem[Condon et al.(1998)]{Con98} Condon J. J. et al., 1998 AJ 115, 1693
\bibitem[Croft et al.(2006)]{Crof06} Croft S. et al., 2006, ApJ, 647, 1040
\bibitem[Croston et al.(2005)]{Cro05} Croston J. H., Hardcastle M. J., Harris D. E., Belsole E., Birkinshaw M., Worrall D. M., 2005, ApJ, 626, 733
\bibitem[Croton et al.(2006)]{Cro:06} Croton D. J. et al., 2006, MNRAS, 365, 11
\bibitem[Cruz et al.(2006)]{cru06} Cruz M. J. et al., 2006, MNRAS, 373, 1531.
\bibitem[De Breuck et al.(2000a)]{deB00a}  De Breuck C., van Breugel W., R{\"o}ttgering H. J. A., Miley G., 2000a, A\&AS, 143, 303
\bibitem[De Breuck et al.(2000b)]{deB00b}  De Breuck C., R{\"o}ttgering H. J. A., Miley G., van Breugel W., Best P., 2000b, A\&A, 362, 519
\bibitem[De Breuck et al.(2001)]{deB01}  De Breuck C., et al., 2001, AJ, 121, 1241
\bibitem[De Breuck et al.(2002)]{deB02} De Breuck C., van Breugel W., Stanford S. A., R{\"o}ttgering H. J. A., Miley G., Stern D., 2002, AJ, 123, 637. 
\bibitem[De Breuck et al.(2004)]{deB04} De Breuck C., Hunstead R. W., Sadler E. M., Rocca-Volmerange B., Klamer I., 2004, MNRAS 347, 837.
\bibitem[De Breuck et al.(2006)]{deB06} De Breuck, Klamer I., Johnston H., Hunstead R. W., Bryant J. J., Rocca-Volmerange B., Sadler E. M., 2006, MNRAS, 366, 58.
\bibitem[De Lucia et al.(2005)]{deL:05}  De Lucia G., Springel V., White S. D. M., Croton D., Kauffmann G., 2006, MNRAS, 366, 499
\bibitem[Dekker, Delabre \& D'Odorico(1986)]{Dek86} Dekker H., Delabre B., D'Odorico S.,  1986, in SPIE: Instrumentation in astronomy VI; Proceedings of the Meeting, Tucson, AZ, Mar. 4-8, 1986. Part 1, 627, 39
\bibitem[Dopita \& Sutherland(1996)]{Dop96} Dopita M., Sutherland R., 1996, ApJS, 102, 161
\bibitem[Eales et al.(1997)]{eal97} Eales, Rawlings S., Law-green D., Cotter G., Lacy M., 1997, MNRAS, 291, 593.
\bibitem[El Bouchefry \& Cress(2007)]{ElB07} El Bouchefry K., Cress C. M., 2007, AN, 328, 577
\bibitem[Feulner et al.(2005)]{Feu05} Feulner G., Gabasch A., Salvato M., Drory N., Hopp U., Bender R., 2005, ApJ, 633, L9
\bibitem[Fioc \& Rocca-Volmerange(1997)]{fio97} Fioc M., Rocca-Volmerange B., 1997, A\&A, 326, 950.
\bibitem[Gillingham \& Jones(2000)]{gil00} Gillingham P. Jones D., 2000, in Masanori I., Moorwood A. F. M., eds., Proc. SPIE: Optical and IR Telescope Instrumentation and Detectors, 4008, 1084.
\bibitem[Hambly et al.(2001)]{Ham01} Hambly N. C. et al., 2001, MNRAS, 326, 1279.
\bibitem[Helfand et al.(1999)]{Hel99} Helfand D., Schnee S., Becker R., White R., McMahon R., 1999, AJ, 117, 1568
\bibitem[Hobbs(1984)]{Hob84} Hobbs L. M., 1984, ApJ, 280, 132
\bibitem[Hopkins(2004)]{Hop04} Hopkins A. M., 2004, ApJ, 615, 209
\bibitem[Hopkins \& Beacom(2006)]{Hop06} Hopkins A. M., Beacom J. F., 2006, ApJ, 651, 142
\bibitem[Jarvis et al.(2001)]{jar01} Jarvis M. J., Rawling S., Eales S., Blundell K. M., Bunker A. J., Croft S., McLure R. J., Willott C. J., 2001, MNRAS, 326, 1585.
\bibitem[Kapahi, Subrahmanya \& Kulkarni(1987)]{Kap87} Kapahi V. K., Subrahmanya C. R., Kulkarni V.K., 1987 J. Astophys. Astron., 8, 33.
\bibitem[Kellermann, Pauliny-Toth \& Williams(1969)]{Kel69} Kellermann K. I., Pauliny-Toth I. I. K., Williams P. J. S., 1969, ApJ, 157, 1
\bibitem[Kodama et al.(2007)]{Kod07} Kodama T., Tanaka I., Kajisawa M., Kurk J., Venemans B., De Breuck C., Vernet J., Lidman C., 2007, MNRAS, 377, 1717
\bibitem[Klamer et al.(2004)]{Kla:04}  Klamer, I. J., Ekers R. D., Sadler E. M., Hunstead R.W., 2004, ApJ, 612, L97
\bibitem[Klamer et al.(2006)]{kla06} Klamer I., Ekers R. D., Bryant J. J., Hunstead R. W., Sadler E. M., De Breuck C., 2006, MNRAS, 371, 852.
\bibitem[Kurk et al.(2000a)]{Kur00a} Kurk, J. D., R{\"o}ttgering H. J.~A., Pentericci, L., Miley, G. K., 2000a, in Mazure A., Le F\'evre O., Le Brun V., eds, ASP Conf. Ser.: Clustering at high redshift, 200, 424.
\bibitem[Kurk et al.(2000b)]{Kur00b} Kurk, J. D. et al., 2000b, A\&A, 358, L1 
\bibitem[Laing \& Peacock(1980)]{Lai80} Laing R. A., Peacock J. A., 1980, MNRAS, 190, 903
\bibitem[Large et al.(1981)]{Lar81} Large, M. I., Mills B. Y., Little A. G., Crawford D. F., Sutton J. M., 1981 MNRAS, 194, 693
\bibitem[Lehnert et al.(1999)]{Leh99} Lehnert M. D., van Breugel W. J. M., Heckman T. M., Miley G., 1999, ApJS, 124, 11
\bibitem[Lilly et al.(1996)]{Lil96} Lilly S. J., Le F\`{e}vre O., Hammer F., Crampton D., 1996, ApJ, 460, L1
\bibitem[Madau et al.(1996)]{Mad96} Madau P., Ferguson H. C., Dickinson M. E., Giavalisco M., Steidel C. C., Fruchter A., 1996, MNRAS, 283, 1388
\bibitem[Martini et al.(2004)]{mar04} Martini P., Persson S. E., Murphy D. C., Birk C., Schectman S. A., Gunnels S. M., Koch E., 2004, in Moorwood A. F. M., Masanori I., eds, Proc SPIE: Ground-based Instrumentation for Astronomy, 5492, 1653.
\bibitem[Mauch et al.(2003)]{Mau03} Mauch T., Murphy T., Buttery H. J., Curran J., Hunstead R. W., Piestrzynski B., Robertson J. G., Sadler E., 2003 MNRAS 342, 1117
\bibitem[McCarthy(1993)]{McC93} McCarthy P. J., 1993, ARA\&A, 31, 639
\bibitem[McCarthy et al.(1996)]{McC96} McCarthy P. J., Kapahi V. K., van Breugel W., Persson S. E., Athreya R. M., Subrahmanya C. R., 1996, ApJS, 107, 19
\bibitem[Nelson et al.(2001)]{Nel:01} Nelson A. E., Gonzalez A. H., Zaritsky D., Dalcanton J. J., 2001, ApJ, 563, 629
\bibitem[Owen \& Ledlow(1997)]{Owe97} Owen F. N., Ledlow M. J., 1997, ApJS, 108, 41
\bibitem[Pandey (2006)]{Pan:06} Pandey, V., 2006, PhD thesis, Raman Research Institute, Bangalore, India.
\bibitem[Pentericci et al.(2000)]{Pen00} Pentericci L., van Reeven W., Carilli C. L., R{\"o}ttgering H. J.~A., Miley G. K., 2000, A\&AS, 145, 121
\bibitem[Pentericci et al.(1999)]{Pen99} Pentericci L., R{\"o}ttgering H. J.~A., Miley G. K., McCarthy P., Spinrad H., van Breugel W. J. M., Macchetto F., 1999, A\&AS, 341, 329
\bibitem[Prestage \& Peacock(1988)]{Pre88} Prestage R. M., Peacock J. A., 1988, MNRAS, 230, 131
\bibitem[Rees(1989)]{Ree:89}  Rees, M. J., 1989, MNRAS, 239, 1P
\bibitem[Reuland et al.(2007)]{reu:07} Reuland, M. et al. 2007, AJ, 133, 2607.
\bibitem[Riley et al.(1989)]{Ril89} Riley J. M., Warner P. J., Rawling S., Saunders R., Pooley G. G., Eales S. A., 1988, MNRAS, 236, 13p
\bibitem[Rocca-Volmerange et al.(2004)]{Roc04} Rocca-Volmerange B., Le Borgne D., De Breuck C., Fioc M., Moy E., 2004, A\&A, 415, 931.
\bibitem[Rodgers, Conroy \& Bloxham (1988)]{rcb88} Rodgers A.~W.,  Conroy P.,    Bloxham G.,  1988, PASP, 100, 626
\bibitem[R{\"o}ttgering et al.(1997)]{Rott97} R{\"o}ttgering H. J.~A.,  van Ojik R.,  Miley G.~K.,  Chambers K.~C.,  van Breugel W. J.~M., de Koff S., 1997, A\&A, 326, 505
\bibitem[Seymour et al.(2007)]{Sey07} Seymour N. et al., 2007, ApJS 171, 353.
\bibitem[Silk(2005)]{Sil:05}  Silk, J., 2005, MNRAS, 364, 1337
\bibitem[Skrutskie et al.(2006)]{Skr06} Skrutskie M. F. et al., 2006, AJ, 131, 1163
\bibitem[Springel et al.(2005)]{Spr:05}  Springel V., Di Matteo T., Hernquist L., 2005, ApJ, 620, L79
\bibitem[Stanford, Eisenhardt \& Dickinson (1998)]{Sta:98} Stanford S. A., Eisenhardt P. R., Dickinson M., 1998, ApJ, 492, 461
\bibitem[Steidel et al.(2005)]{Ste05} Steidel C. C., Adelberger K. L., Shapley A. E., Erb D., Reddy N., Pettini M., 2005, ApJ, 626, 44
\bibitem[Thomas et al.(2005)]{Tho:05} Thomas D., Maraston C., Bender R., Mendez de Oliveira C., 2005, ApJ, 621, 673
\bibitem[Urry \& Padovani(1995)]{urr95} Urry C.M., Padovani P., 1995, PASP, 107, 803
\bibitem[van Breugel et al.(1998)]{vanB98} van Breugel W. J. M., Stanford S. A., Spinrad H., Stern D., Graham J. R., 1998, ApJ, 502, 614
\bibitem[Vanden Berk et al.(2001)]{Vand01} Vanden Berk D. E. et al., 2001, AJ, 122, 549
\bibitem[van Ojik et al.(1997)]{vanO97} van Ojik R., R{\"o}ttgering H. J.~A., Miley G.~K., Hunstead R. W., 1997, A\&A, 317, 358
\bibitem[Venemans et al.(2002)]{Ven:02}  Venemans, B. P. et al., 2002, ApJ, 569, L11
\bibitem[Venemans et al.(2007)]{Ven:07}  Venemans, B. P. et al., 2007, A\&A, 461, 823
\bibitem[Villar-Mart\'{i}n, Binette \& Fosbury(1996)]{Vil96} Villar-Mart\'{i}n M., Binette L., Fosbury R. A. E., 1996, A\&A, 312, 751
\bibitem[Villar-Mart\'{i}n, Tadhunter \& Clark(1997)]{Vil97} Villar-Mart\'{i}n M., Tadhunter C., Clark N., 1997, A\&A, 323, 21
\bibitem[White \& Becker(1992)]{Whi92} White R. L., Becker R. H., 1992, ApJS, 79, 331
\bibitem[Willott et al.(1999)]{Wil99} Willott C., Rawlings S., Blundell K., Lacy M., 1999, MNRAS, 309, 1017
\bibitem[Willott et al.(2003)]{Wil03} Willott C., Rawlings S., Jarvis M., Blundell K., 2003, MNRAS, 339, 173

\end{thebibliography}
\end{document}